\documentclass[useAMS,usenatbib]{mn2e}

\usepackage{txfonts}
\usepackage{amssymb}
\usepackage{graphicx}
\usepackage{subfigure}
\usepackage{url}

\usepackage{ifpdf}
\ifpdf
  \usepackage[pdftex,
            pdfauthor={Jason McEwen},
            pdftitle={Non-Gaussianity in the Bianchi VIIh corrected WMAP 1-year data},
            pdfsubject={CMB Non-Gaussianity},
            pdfkeywords={cosmic microwave background -- methods: data analysis -- methods: numerical},
            pdfproducer={pdflatex with hyperref},
            pdfcreator={pdflatex}]{hyperref}
\else
  \usepackage[dvips,hypertex]{hyperref}
\fi

\newcommand{\eqn}[1]
	{(#1)}

\newcommand{\tbl}[1]
	{Table~#1}

\newcommand{\fig}[1]
	{Fig.~#1}

\newcommand{\sectn}[1]
	{section~#1}

\newcommand{\appn}[1]
	{appendix~#1}

\newcommand{\bianchi}
	{{Bianchi}}
\newcommand{\bianchiviih}
	{{Bianchi VII$_{\rm h}$}}

\newcommand{\cmb}
	{{CMB}}
\newcommand{\cmbtext}
	{cosmic microwave background}

\newcommand{\cswt}
	{{CSWT}}
\newcommand{\cswttext}
	{continuous spherical wavelet transform}

\newcommand{\mexhat}
	{Mexican hat}
\newcommand{\Mexhat}
	{Mexican hat}
\newcommand{\morlet}
	{real Morlet}
\newcommand{\Morlet}
	{Real Morlet}

\newcommand{\wmap}
        {{WMAP}}
\newcommand{\wmapbianchi}
        {{WMAP}$_{\rm{B.VII_h}}$}
\newcommand{\wmaptext}
        {Wilkinson Microwave Anisotropy Probe}
\newcommand{\ilctext}
	{{internal linear combination}}
\newcommand{\ilc}
	{{ILC}}
\newcommand{\cobe}
	{\mbox{COBE-DMR}}
\newcommand{\cobetext}
        {Cosmic Background Explorer-Differential Microwave Radiometer}
\newcommand{\lcdm}
	{$\Lambda${CDM}}
\newcommand{\lcdmtext}
	{Lambda Cold Dark Matter}
\newcommand{\lambdaarch}
	{{LAMBDA}}

\newcommand{\mnras}{\mbox{MNRAS}}
\newcommand{\physd}{Phys. Rev. D.}
\newcommand{\physlett}{Phys. Rev. Lett.}
\newcommand{\apj}{ApJ}
\newcommand{\apjl}{ApJL}
\newcommand{\apjs}{ApJS}
\newcommand{\acha}{Applied and Computational Harmonic Analysis}
\newcommand{\jmp}{J.\ of Math.\ Phys.}

\newcommand{\kpzero}
	{{Kp0}}

\newcommand{\healpix}
	{{HEALPix}}

\newcommand{\etal}
	{\mbox{et al.}}

\newcommand{\ie}
	{\mbox{i.e.}}

\newcommand{\jaffeshort}{{J05}}

\newcommand{\el}{\ensuremath{\ell}}
\newcommand{\m}{\ensuremath{m}}

\newcommand{\sh}[3]{\ensuremath{Y_{#1#2}({#3})}}
\newcommand{\shc}[3]{\ensuremath{Y_{#1#2}^{\conj}({#3})}}

\newcommand{\aleg}[3]{\ensuremath{P_{#1}^{#2}({#3})}}
\newcommand{\alm}{\ensuremath{a}}
\newcommand{\almi}{\ensuremath{a_{\el\m}}}
\newcommand{\almpi}{\ensuremath{a_{\el\m\p}}}
\newcommand{\almitilde}{\ensuremath{\tilde{a}_{\el\m}}}
\newcommand{\domega}{\ensuremath{{\rm d}\Omega}}

\newcommand{\bx}{\ensuremath{x}}
\newcommand{\bhand}{\ensuremath{\kappa}}
\newcommand{\bh}{\ensuremath{h}}
\newcommand{\omegat}{\ensuremath{\Omega_0}}
\newcommand{\ze}{\ensuremath{z_E}}
\newcommand{\hub}{\ensuremath{H}}
\newcommand{\bshear}{\ensuremath{\left(\frac{\sigma}{H}\right)_0}}
\newcommand{\bvort}{\ensuremath{\left(\frac{\omega}{H}\right)_0}}
\newcommand{\ba}{\ensuremath{A}}
\newcommand{\bb}{\ensuremath{B}}
\newcommand{\bi}[2]{\ensuremath{I^{#1}_{#2}}}
\newcommand{\thetaph}{\ensuremath{\theta_0}}
\newcommand{\phiph}{\ensuremath{\phi_0}}
\newcommand{\thetaob}{\ensuremath{\theta_{\rm ob}}}
\newcommand{\phiob}{\ensuremath{\phi_{\rm ob}}}
\newcommand{\thetaalm}{\ensuremath{\theta}}
\newcommand{\phialm}{\ensuremath{\phi}}
\newcommand{\bs}{\ensuremath{s}}
\newcommand{\bt}{\ensuremath{\tau}}
\newcommand{\bps}{\ensuremath{\psi}}
\newcommand{\bcone}{\ensuremath{C_1}}
\newcommand{\bctwo}{\ensuremath{C_2}}
\newcommand{\bcthree}{\ensuremath{C_3}}
\newcommand{\dtemp}{\ensuremath{\Delta T}}
\newcommand{\temp}{\ensuremath{T_0}}
\newcommand{\kron}{\ensuremath{\delta}}

\renewcommand{\exp}[1]{\ensuremath{{\rm e}^{#1}}}
\newcommand{\img}
	{\ensuremath{\mathit{i}}}
\newcommand{\dx}
        {\ensuremath{\mathrm{\,d}}}

\newcommand{\scale}
	{\ensuremath{a}}
\newcommand{\effsize}
	{\ensuremath{\xi}}

\newcommand{\eccen}
	{\ensuremath{\epsilon}}

\newcommand{\eulera}
	{\ensuremath{\alpha}}
\newcommand{\eulerb}
	{\ensuremath{\beta}}
\newcommand{\eulerc}
	{\ensuremath{\gamma}}
\newcommand{\eulers}
	{\ensuremath{\eulera, \eulerb, \eulerc}}

\newcommand{\lmax}
	{\ensuremath{\el_{\rm max}}}

\newcommand{\conj}
	{\ensuremath{\ast}}

\newcommand{\sphere}
	{\ensuremath{{S^2}}}

\newcommand{\num}
	{\ensuremath{N}}
\newcommand{\p}
	{\ensuremath{^\prime}}

\newcommand{\spcend}
	{\ensuremath{\:}}

\newcommand{\nstd}
	{\ensuremath{\num_\sigma}}
\newcommand{\ndev}
	{\ensuremath{\num_{\rm dev}}}

\newcommand{\conflevel}
	{\ensuremath{\delta}}

\newcommand{\spotloc}{\mbox{\ensuremath{(l,b)=(209^\circ,-57^\circ)}}}

\newcommand{\ngsim}
	{1000}

\newcommand{\nstatmexskew}
	{28}
\newcommand{\nstatmexkurt}
	{47}

\newcommand{\nstdmexskewsgn}
	{\mbox{$-3.38$}}
\newcommand{\nstdmexkurtsgn}
	{\mbox{$3.12$}}
\newcommand{\clmexskew}
	{97.2}
\newcommand{\clmexkurt}
	{95.3}

\newcommand{\nstatmexepskew}
	{39}
\newcommand{\nstatmexepkurt}
	{199}

\newcommand{\nstdmexepskewsgn}
	{\mbox{$-4.10$}}
\newcommand{\nstdmexepkurtsgn}
	{\mbox{$3.01$}}
\newcommand{\clmexepskew}
	{96.1}
\newcommand{\clmexepkurt}
	{80.1}

\newcommand{\nstatmorskew}
	{17}
\newcommand{\nstatmorkurt}
	{642}

\newcommand{\nstdmorskewteg}  
	{\mbox{$6.42$}}

\newcommand{\nstdmorskewsgn}
	{\mbox{$-5.61$}}
\newcommand{\nstdmorskewtegsgn}  
	{\mbox{$-6.42$}}
\newcommand{\nstdmorkurtsgn}
	{\mbox{$2.66$}}
\newcommand{\clmorskew}
	{98.3}
\newcommand{\clmorkurt}
	{35.8}

\newcommand{\nstatmexskewbian}{21}
\newcommand{\nstatmexkurtbian}{605}

\newcommand{\nstdmexskewsgnbian}{\mbox{$-3.53$}}
\newcommand{\nstdmexkurtsgnbian}{\mbox{$1.70$}}
\newcommand{\clmexskewbian}{97.9}
\newcommand{\clmexkurtbian}{39.5}

\newcommand{\nstatmexepskewbian}{29}
\newcommand{\nstatmexepkurtbian}{887}

\newcommand{\nstdmexepskewsgnbian}{\mbox{$-4.25$}}
\newcommand{\nstdmexepkurtsgnbian}{\mbox{$1.88$}}
\newcommand{\clmexepskewbian}{97.1}
\newcommand{\clmexepkurtbian}{11.3}

\newcommand{\nstatmorskewbian}{16}
\newcommand{\nstatmorkurtbian}{628}

\newcommand{\nstdmorskewsgnbian}{\mbox{$-5.66$}}
\newcommand{\nstdmorkurtsgnbian}{\mbox{$2.67$}}
\newcommand{\clmorskewbian}{98.4}
\newcommand{\clmorkurtbian}{37.2}

\title[Non-Gaussianity in the \wmap\ 1-year data]
  {Non-Gaussianity detections in the \bianchiviih\ corrected \wmap\ 1-year %
   data made with directional spherical wavelets}

\author[J.~D.~McEwen \etal]
  {J.~D.~McEwen$^1$\thanks{E-mail: mcewen@mrao.cam.ac.uk},
   M.~P.~Hobson$^1$, A.~N.~Lasenby$^1$ and D.~J.~Mortlock$^{2,3}$\\
  $^1$Astrophysics Group, 
      Cavendish Laboratory, Madingley Road,
      Cambridge CB3 0HE, UK\\
  $^2$Institute of Astronomy, Madingley Road,
      Cambridge CB3 0HA, UK\\
  $^3$Blackett Laboratory, Imperial College of Science, Technology and Medicine,
    Prince Consort Road, London SW7 2BW, UK}
\date{\today}
\pagerange{\pageref{firstpage}--\pageref{lastpage}} 
\pubyear{2005}

\def\LaTeX{L\kern-.36em\raise.3ex\hbox{a}\kern-.15em
    T\kern-.1667em\lower.7ex\hbox{E}\kern-.125emX}

\begin{document}
\label{firstpage}
\maketitle


\begin{abstract}
Many of the current anomalies reported in the \wmaptext\ (\wmap) 1-year data disappear after `correcting' for the best-fit embedded Bianchi type VII$_{\rm h}$ component \citep{jaffe:2005}, albeit assuming no dark energy component.
%
We investigate the effect of this Bianchi correction on the detections of non-Gaussianity in the \wmap\ data that we previously made using directional spherical wavelets \citep{mcewen:2005a}.  
We confirm that the deviations from Gaussianity in the kurtosis of spherical \mexhat\ wavelet \mbox{coefficients} are eliminated once the data is corrected for the Bianchi component, as previously discovered by \citet{jaffe:2005}.
This is due to the reduction of the cold spot at Galactic coordinates \spotloc, which \citet{cruz:2005} claim to be the sole source of non-Gaussianity introduced in the kurtosis.  Our previous detections of non-Gaussianity observed in the skewness of spherical wavelet coefficients are not reduced by the Bianchi correction.  Indeed, the most significant detection of non-Gaussianity made with the spherical \morlet\ wavelet at a significant level of \clmorskewbian\% remains (using a very conservative method to estimate the significance).  Furthermore, we perform preliminary tests to determine if foregrounds or systematics are the source of this non-Gaussian signal, concluding that it is unlikely that these factors are responsible.  We make our code to simulate Bianchi induced temperature fluctuations publicly available.
%
\end{abstract}




\section{Introduction}

Recent measurements of the \cmbtext\ (\cmb) anisotropies made by the \wmaptext\ (\wmap)
provide full-sky data of unprecedented precision on which to test the standard cosmological model.  One of the most important and topical assumptions of the standard model currently under examination is that of the statistics of the primordial fluctuations that give rise to the anisotropies of the \cmb.
Recently, the assumption of Gaussianity has been questioned with many works highlighting deviations from Gaussianity in the \wmap\ 1-year data.

A wide range of Gaussianity analyses have been performed 
on the \wmap\ 1-year data,
calculating measures such as
the bispectrum and Minkowski functionals
  \citep{komatsu:2003,mm:2004,lm:2004},
the genus
  \citep{cg:2003,eriksen:2004}, 
correlation functions
  \citep{gw:2003,eriksen:2005,tojeiro:2005},
low-multipole alignment statistics 
  \citep{oliveira:2004,copi:2004,copi:2005,schwarz:2004,slosar:2004,weeks:2004,lm:2005a,lm:2005b,lm:2005c,lm:2005d,bielewicz:2005}, 
phase associations
  \citep{chiang:2003,coles:2004,dineen:2005},
local curvature 
  \citep{hansen:2004,cabella:2005},
the higher criticism statistic
  \citep{cayon:2005},
hot and cold spot statistics
  \citep{larson:2004,larson:2005,cruz:2005}
and wavelet coefficient statistics 
  \citep{vielva:2003,mw:2004,mcewen:2005a}.
Some statistics show consistency with Gaussianity, whereas others provide
some evidence for a non-Gaussian signal and/or an asymmetry
between the northern and southern Galactic hemispheres.  Although these detections may simply highlight unremoved foreground contamination or other systematics in the \wmap\ data, it is important to also consider non-standard cosmological models that could give rise to non-Gaussianity.

One such alternative is that the universe has a small universal shear and rotation -- these are the so-called Bianchi models.
Relaxing the assumption of isotropy about each point yields more complicated solutions to Einstein's field equations that contain the 
Friedmann-Robertson-Walker metric as a special case.
\citet{barrow:1985} derive the induced \cmb\ temperature fluctuations that result in the Bianchi models, however they do not include any dark energy component as it was not considered plausible at the time.
There is thus a need for new derivations of solutions to the Bianchi models in a more modern setting.  Nevertheless, the induced \cmb\ temperature fluctuations derived by \citet{barrow:1985} provide a good phenomenological setting in which to examine and raise the awareness of more exotic cosmological models.

Bianchi type VII$_{\rm h}$ models have previously been compared both to the \cobetext\ (\cobe) \citep{kogut:1997} and \wmap\ (\citealt{jaffe:2005}; henceforth referred to as \jaffeshort) data to place limits on the global rotation and shear of the universe.
Moreover, \jaffeshort\ find a statistically significant correlation between one of the \bianchiviih\ models and the \wmap\ \ilctext\ (\ilc) map.
They then `correct' the \ilc\ map using the best-fit Bianchi template and, remarkably, find that many of the reported anomalies in the \wmap\ data disappear.
More recently \citet{lm:2005f} perform a modified template fitting technique and, although they do not report a statistically significant template fit, their corrected \wmap\ data is also free of large scale anomalies.

In this paper we are interested to determine if our previous detections of non-Gaussianity made using directional spherical wavelets \citep{mcewen:2005a} are also eliminated when the \wmap\ data is corrected for the best-fit \bianchiviih\ template determined by \jaffeshort.
In \sectn{{\ref{sec:analysis}}} the best-fit Bianchi template embedded in the \wmap\ data is described and used to correct the data, before a brief review of the analysis procedure is given.
Results are presented and discussed in \sectn{\ref{sec:results}}.  Concluding remarks are made in \sectn{\ref{sec:conclusions}}.

\section{Non-Gaussianity analysis}
\label{sec:analysis}

We recently made significant detections of non-Gaussianity using directional spherical wavelets \citep{mcewen:2005a} and are interested to see if these detections disappear when the data are corrected for an embedded Bianchi component.
The best-fit Bianchi template and the correction of the data is described in this section, before we review the analysis procedure.
We essentially repeat the analysis performed by \citet{mcewen:2005a} for the Bianchi corrected maps, hence we do not describe the analysis procedure in any detail here but give only a very brief overview.

\subsection{Bianchi VII$_{\rm \lowercase{h}}$ template}
\label{sec:bianchi_tmpl}

\begin{figure}
\begin{center}
\includegraphics[clip=,angle=0]{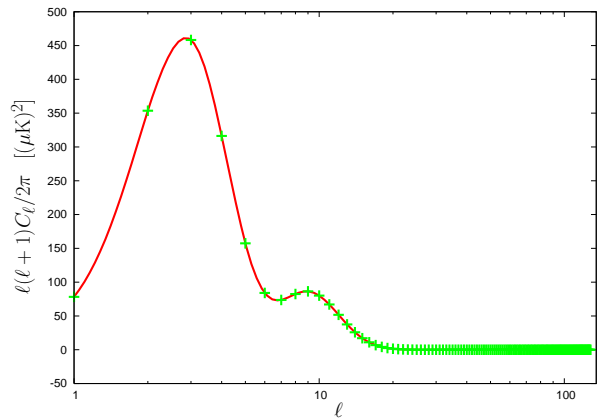}
\caption{Angular power spectrum of the Bianchi-induced temperature fluctuations.  The particular spectrum shown is for the best-fit Bianchi template matched to the \wmap\ data.
Notice that the majority of the power is contained in multipoles below  $\el\sim20$.}
\label{fig:bianchi_cl}
\end{center}
\end{figure}

We have implemented simulations to compute the Bianchi-induced temperature fluctuations, concentrating on the
Bianchi type VII$_{\rm h}$ models, which include the types I, V and VII$_{\rm o}$ as special cases \citep{barrow:1985}.\footnote{Our code to produce simulations of Bianchi type VII-induced temperature fluctuations may be found at: \url{http://www.mrao.cam.ac.uk/~jdm57/}}
Note that the Bianchi type VII$_{\rm h}$ models apply to open or flat universes only.
In \appn{\ref{sec:appn_bianchi}} we describe the equations implemented in our simulation; in particular, we give the analytic forms for directly computing Bianchi-induced temperature fluctuations in both real and harmonic space in sufficient detail to reproduce our simulated maps.  The angular power spectrum of a typical Bianchi-induced temperature fluctuation map is illustrated in \fig{\ref{fig:bianchi_cl}} (note that the Bianchi maps are deterministic and anisotropic, hence they are not fully described by their power spectrum).  An example of the swirl pattern typical of Bianchi-induced temperature fluctuations may be seen in \fig{\ref{fig:maps}~(a)} (this is in fact the map that has the power spectrum shown in \fig{\ref{fig:bianchi_cl}}).
Notice that the Bianchi maps have a particularly low band-limit, both globally and azimuthally (\ie\ in both \el\ and \m\ in spherical harmonic space; indeed, only those harmonic coefficients with $\m=\pm1$ are non-zero).

The best-fit Bianchi template that we use to correct the \wmap\ data is simulated with the parameters determined by \jaffeshort\ using the latest shear and vorticity estimates (\jaffeshort; private communication).  This map is illustrated in \fig{\ref{fig:maps}~(d)}.
In our previous non-Gaussianity analysis \citep{mcewen:2005a} we considered the co-added \wmap\ map \citep{komatsu:2003}.  However, the template fitting technique performed by \jaffeshort\ is only straightforward when considering full-sky coverage.
The Bianchi template is therefore matched to the full-sky \ilc\ map since the co-added \wmap\ map requires a galactic cut.  Nevertheless, we consider both the \ilc\ and co-added \wmap\ map hereafter, using the Bianchi template matched to the \ilc\ map to correct both maps.  The Bianchi template and the original and corrected \wmap\ maps that we consider are illustrated in \fig{\ref{fig:maps}}.

\newlength{\mapplotwidth}
\setlength{\mapplotwidth}{55mm}

\begin{figure*}
\begin{minipage}{\textwidth}
\centering
\mbox{
\subfigure[Best-fit Bianchi template (scaled by four) rotated to the Galactic centre for illustration]
  {\includegraphics[clip=,width=\mapplotwidth]{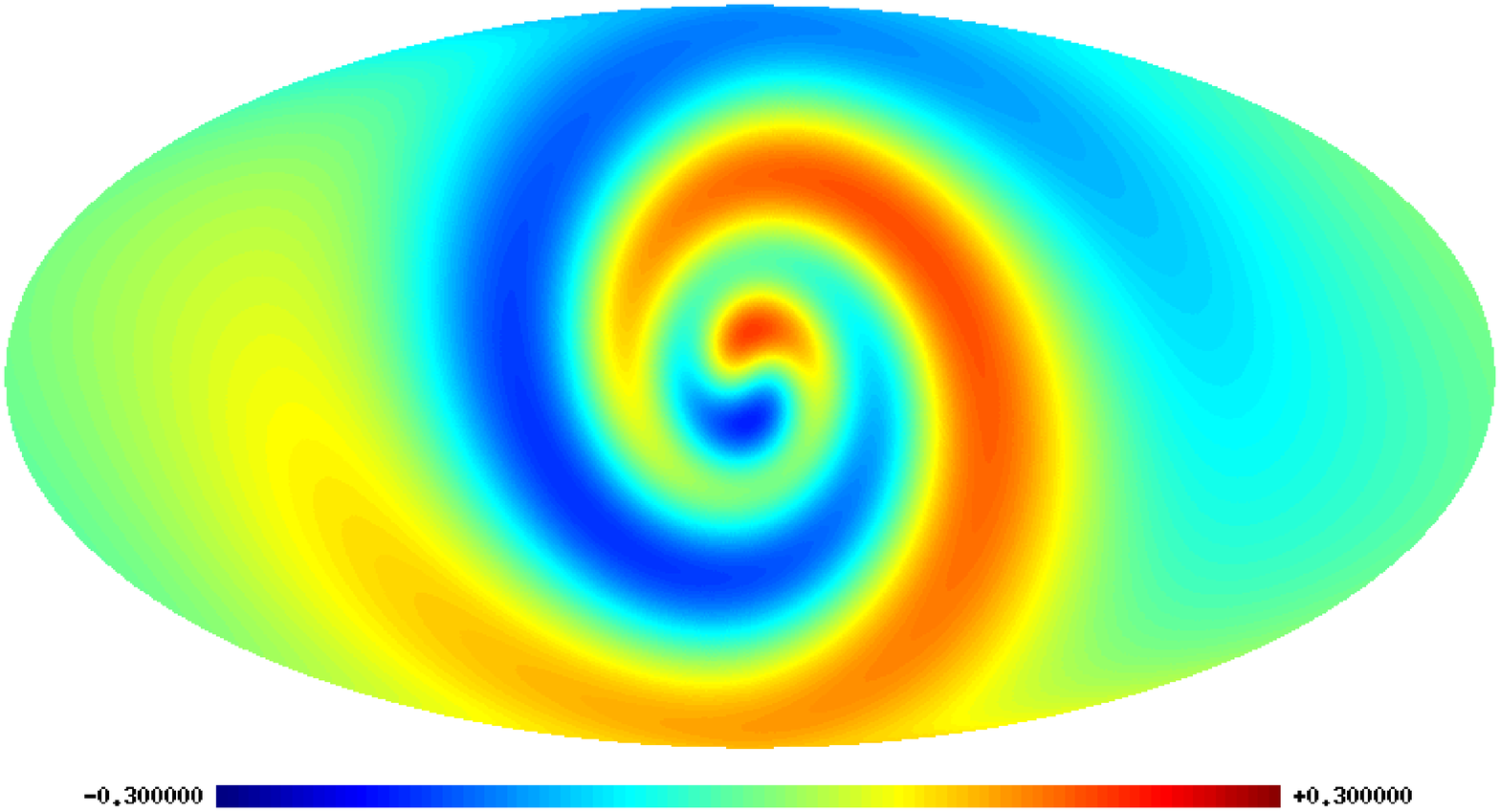}} \quad
\subfigure[\ilc\ map]
  {\includegraphics[clip=,width=\mapplotwidth]{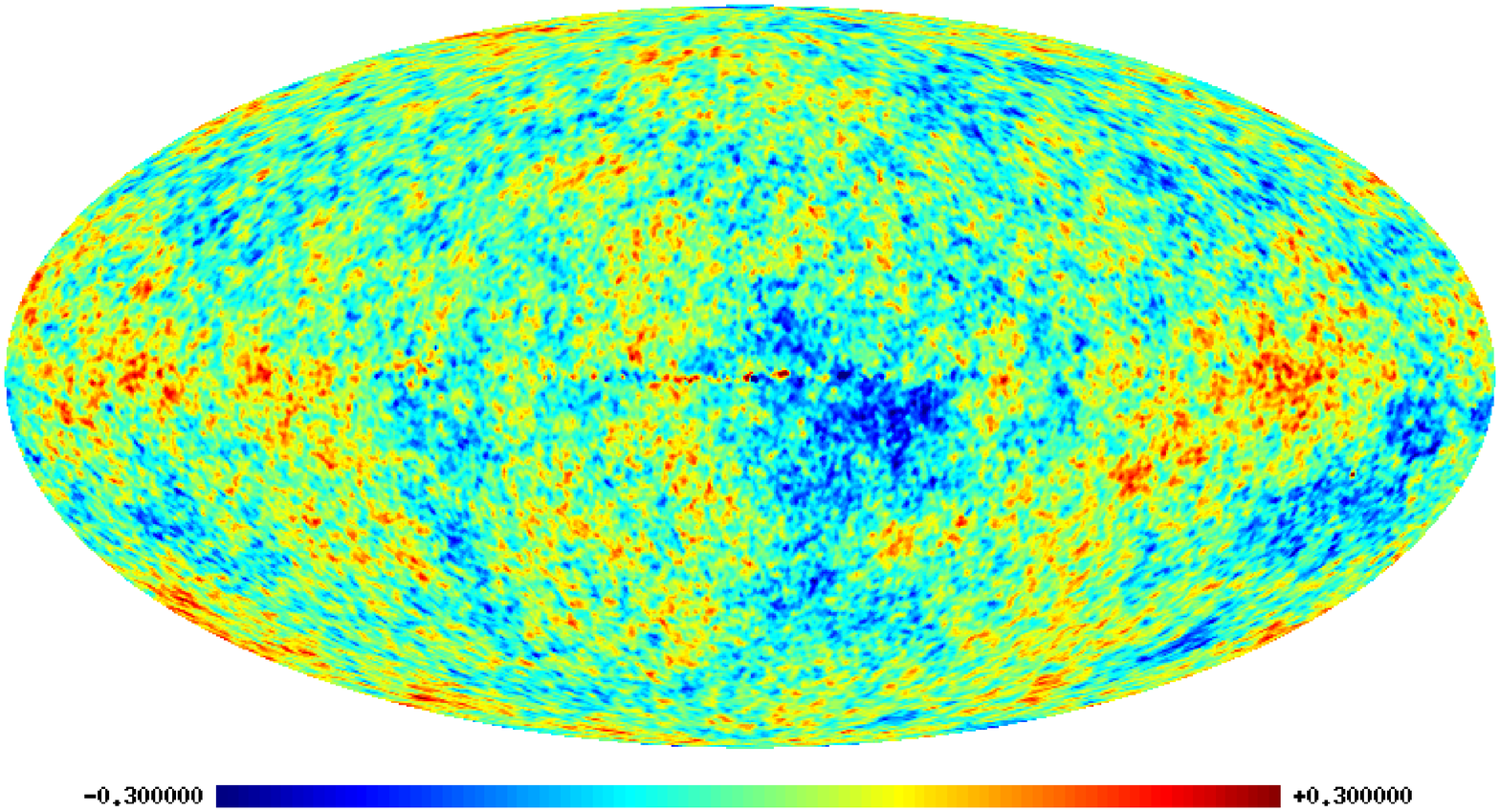}} \quad
\subfigure[\wmap\ co-added map (masked)]
  {\includegraphics[clip=,width=\mapplotwidth]{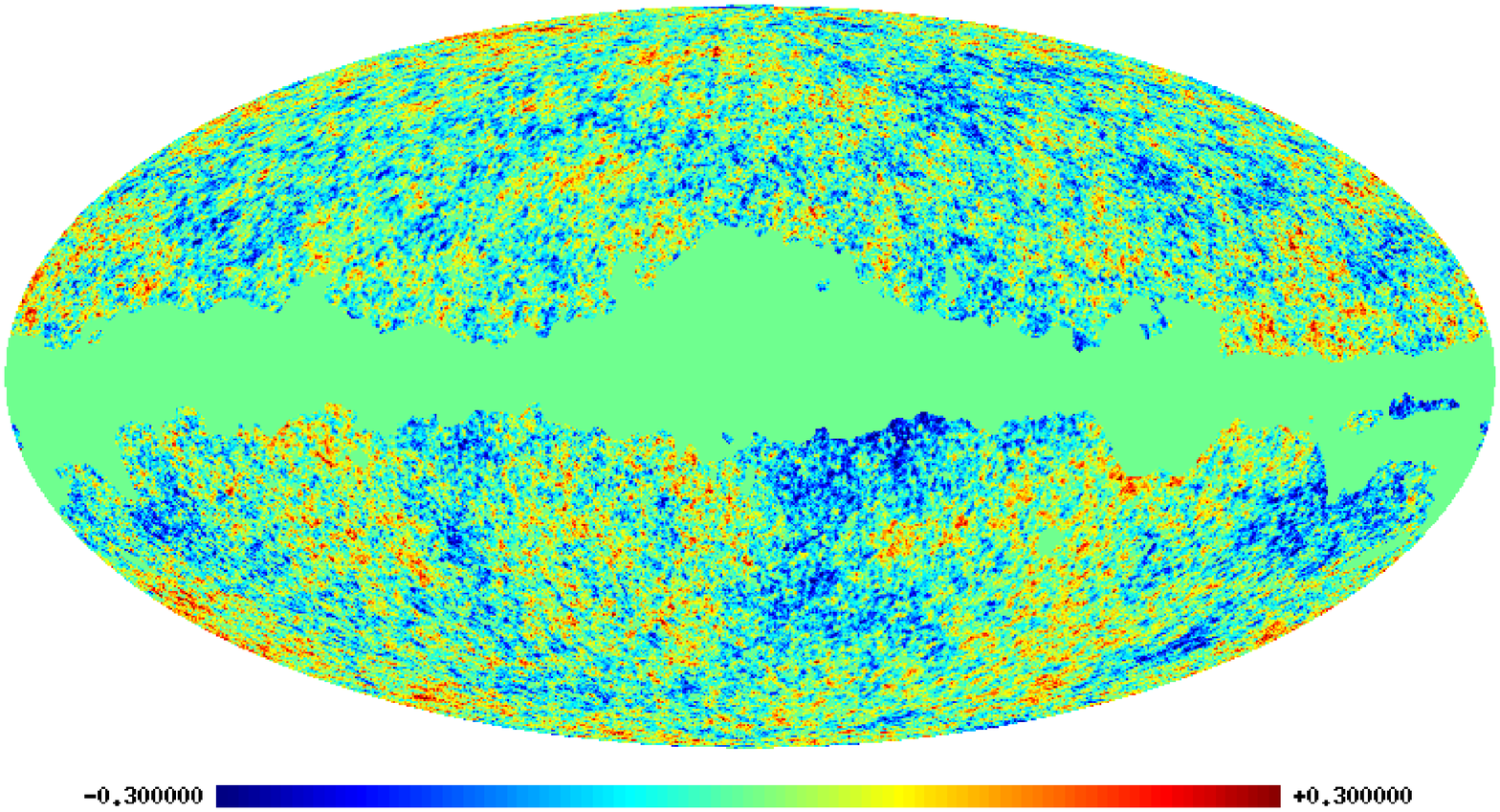}}
}
\mbox{
\subfigure[Best-fit Bianchi template (scaled by four)]
  {\includegraphics[clip=,width=\mapplotwidth]{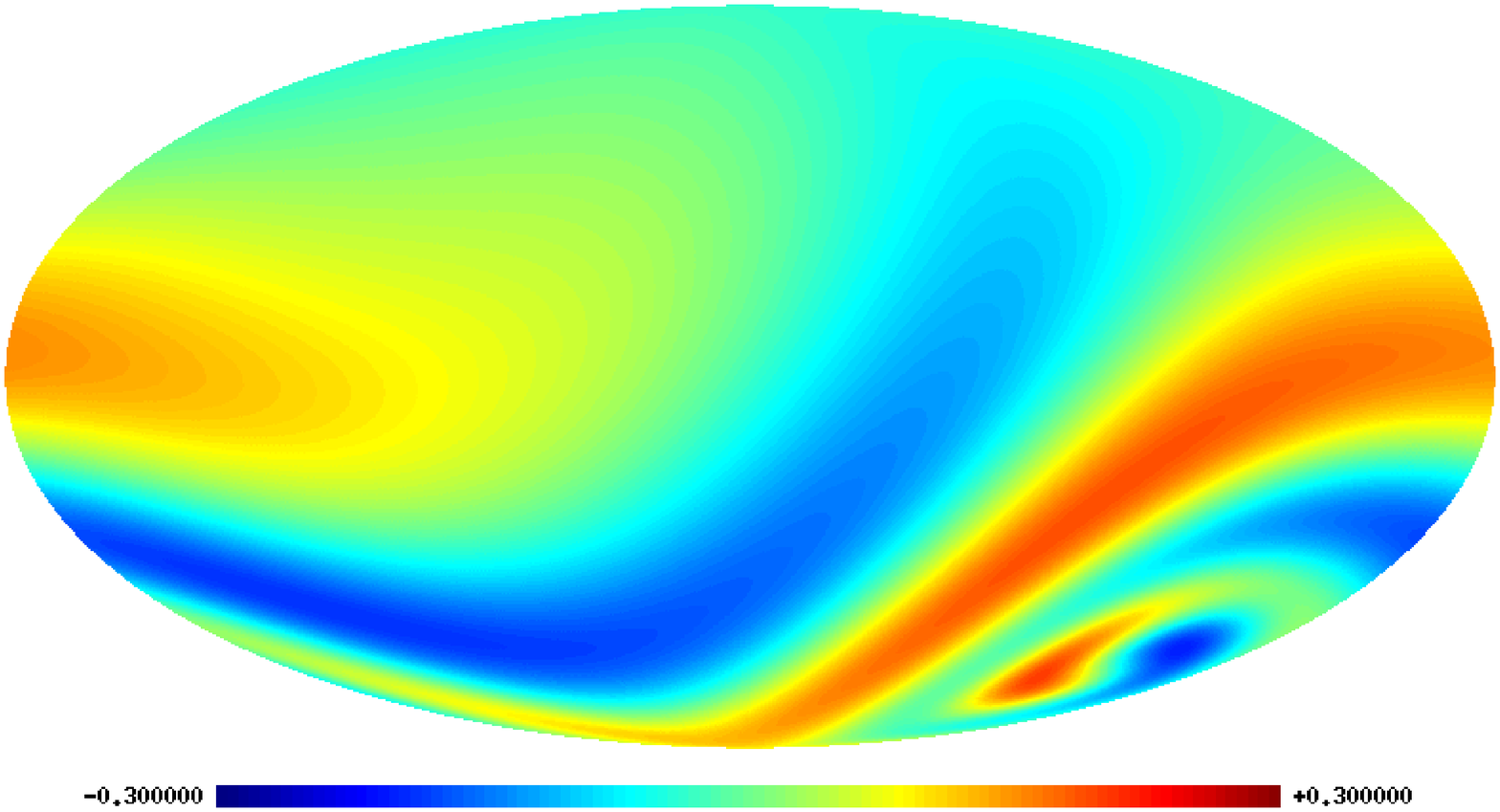}} \quad
\subfigure[Bianchi corrected \ilc\ map]
  {\includegraphics[clip=,width=\mapplotwidth]{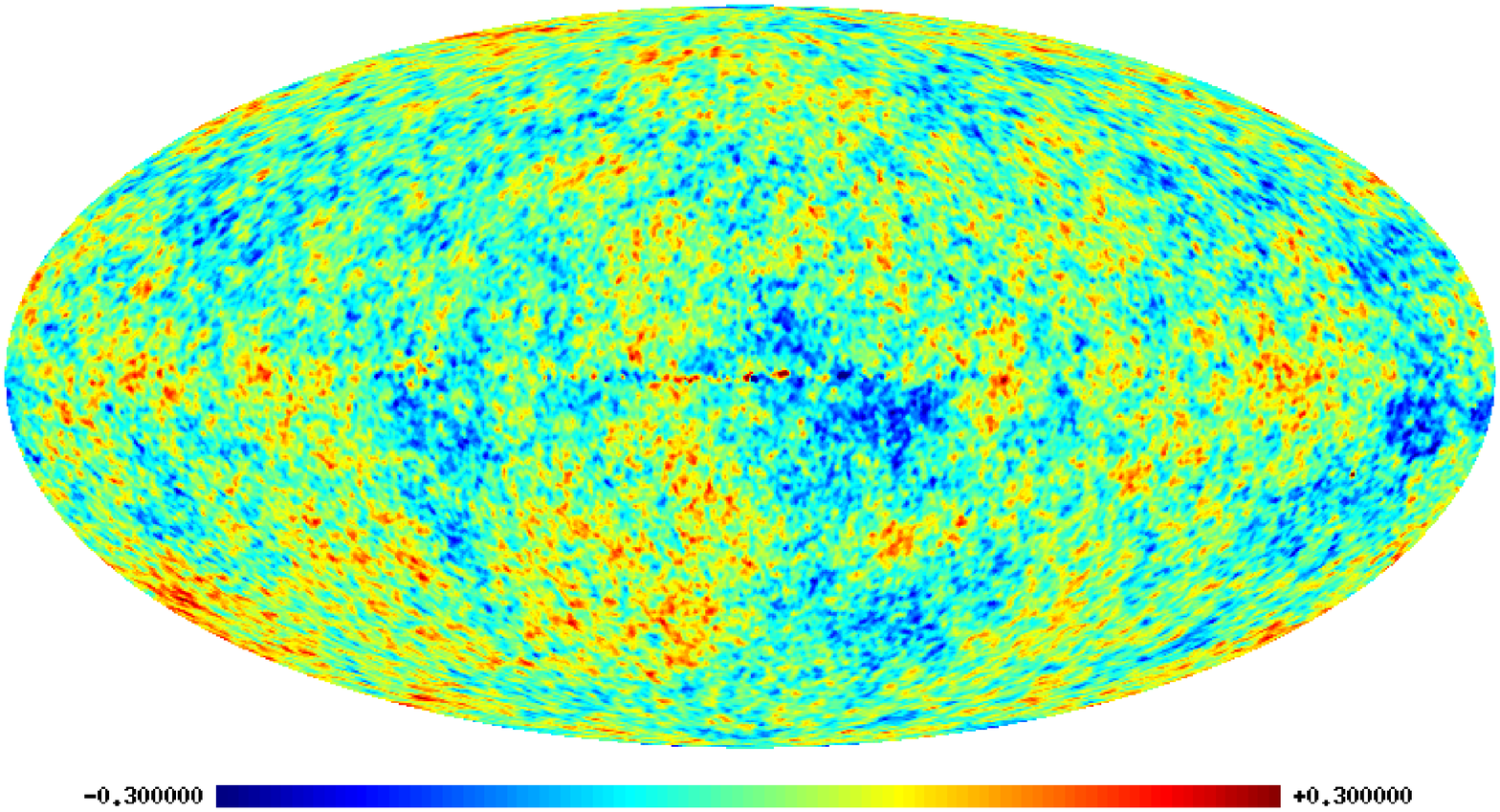}} \quad
\subfigure[Bianchi corrected \wmap\ co-added map (masked)]
  {\includegraphics[clip=,width=\mapplotwidth]{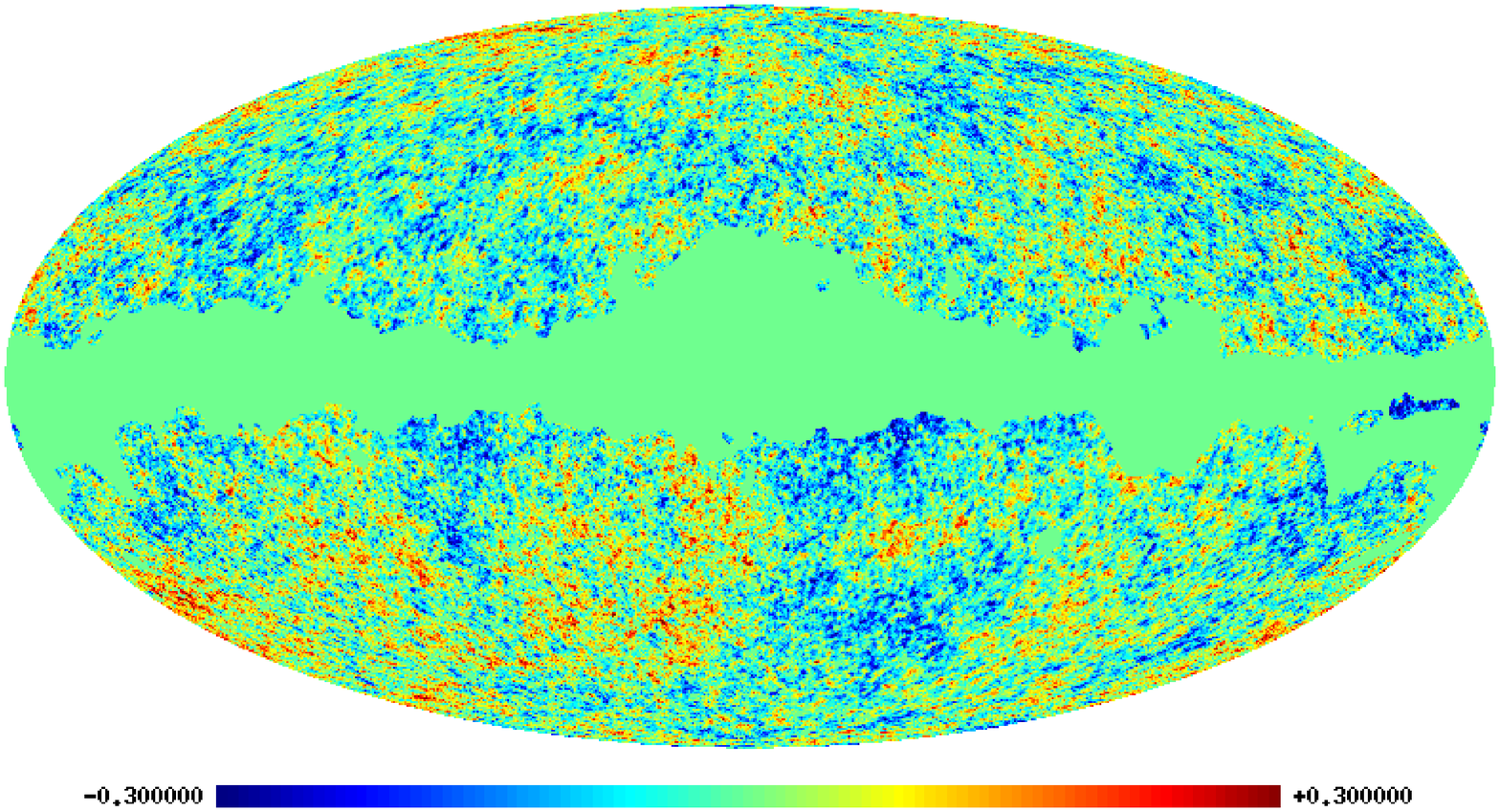}}
}
\caption{Bianchi template and \cmb\ data maps (in mK).  The Bianchi maps are scaled by a factor of four so that the structure may be observed.  The \kpzero\ mask has been applied to the co-added \wmap\ maps.}
\label{fig:maps}
\end{minipage}
\end{figure*}

\subsection{Procedure}

Wavelet analysis is an ideal tool for searching for non-Gaussianity since it allows one to resolve signal components in both scale and space.
The wavelet transform is a linear operation, hence the wavelet coefficients of a Gaussian map will also follow a Gaussian distribution.  One may therefore probe a signal for non-Gaussianity simply by looking for deviations from Gaussianity in the distribution of the wavelet coefficients.

To perform a wavelet analysis of full-sky \cmb\ maps we apply our fast \cswttext\ (\cswt) \citep{mcewen:2005b}, which is based on the spherical wavelet transform developed by Antoine, Vandergheynst and colleagues \citep{antoine:1998,antoine:1999,antoine:2002,antoine:2004,wiaux:2005,wiaux:2005b,wiaux:2005c} and the fast spherical convolution developed by \citet{wandelt:2001}.  In particular, we use the symmetric and elliptical \mexhat\ and \morlet\ spherical wavelets at the scales defined in \tbl{\ref{tbl:scales}}.  The elliptical \mexhat\ and \morlet\ spherical wavelets are directional and so allow one to probe oriented structure in the data.  For the directional wavelets we consider five evenly spaced azimuthal orientations between $[0,\pi)$.
We look for deviations from zero in the skewness and excess kurtosis of spherical wavelet coefficients to signal the presence of non-Gaussianity.
To provide confidence bounds on any detections made, 1000 Monte Carlo simulations are performed on Gaussian \cmb\ realisations produced from the theoretical power spectrum fitted by the \wmap\ team.\footnote{We use the theoretical power spectrum of the \lcdmtext\ (\lcdm) model which best fits the \wmap, Cosmic Background Imager ({CBI}) and Arcminute Cosmology Bolometer Array Receiver ({ACBAR}) \cmb\ data.}
The \ilc\ map, the foreground corrected \wmap\ maps required to create the co-added map, the masks and power spectrum may all be downloaded from the Legacy Archive for Microwave Background Data Analysis (\lambdaarch) website\footnote{\url{http://cmbdata.gsfc.nasa.gov/}}.

\begin{table*}
\begin{minipage}{145mm}
\centering
\caption{Wavelet scales considered in the non-Gaussianity analysis.
  The overall size on the sky $\effsize_1$ for a given
  scale are the same for both the \mexhat\ and \morlet\ wavelets.
  The size on the sky of the internal structure of the \morlet\
  wavelet $\effsize_2$ is also quoted.
}
\label{tbl:scales}
\begin{tabular}{lcccccccccccc} \hline
Scale & 1 & 2 & 3 & 4 & 5 & 6 & 7 & 8 & 9 & 10 & 11 & 12 \\ \hline
Dilation \scale  & 50\arcmin & 100\arcmin & 150\arcmin & 200\arcmin & 250\arcmin & 300\arcmin & 350\arcmin & 400\arcmin & 450\arcmin & 500\arcmin & 550\arcmin & 600\arcmin \\
Size on sky $\effsize_1$ & 141\arcmin & 282\arcmin & 424\arcmin & 565\arcmin & 706\arcmin & 847\arcmin & 988\arcmin & 1130\arcmin & 1270\arcmin & 1410\arcmin & 1550\arcmin & 1690\arcmin \\ 
Size on sky $\effsize_2$ & 15.7\arcmin & 31.4\arcmin & 47.1\arcmin & 62.8\arcmin & 78.5\arcmin & 94.2\arcmin & 110\arcmin & 126\arcmin & 141\arcmin & 157\arcmin & 173\arcmin & 188\arcmin \\ 
\hline
\end{tabular}
\end{minipage}
\end{table*}


\section{Results and discussion}
\label{sec:results}

We examine the skewness and excess kurtosis of spherical wavelet coefficients of the original and Bianchi corrected \wmap\ data to search for deviations from Gaussianity.  Raw statistics with corresponding confidence regions are presented and discussed first, before we consider the statistical significance of detections of non-Gaussianity in more detail.  Localised regions that are the most likely sources of non-Gaussianity are then examined.  Finally, we investigate the possibility of foreground contamination and systematics.

\subsection{Wavelet coefficient statistics}

For a given wavelet, the skewness and kurtosis of wavelet coefficients is calculated for each scale and orientation, for each of the data maps considered.  These statistics are displayed in \fig{\ref{fig:stat_plot}}, with confidence intervals {con\-structed} from the Monte Carlo simulations also shown.  For directional wavelets, only the orientations corresponding to the maximum deviations from Gaussianity are shown.

The significant deviation from Gaussianity previously observed by \citet{vielva:2003}, \citet{mw:2004} and \citet{mcewen:2005a} in the kurtosis of the \mexhat\ wavelet coefficients is reduced when the data are corrected for the Bianchi template, confirming the results of \jaffeshort.  However, it appears that a new non-Gaussian signal may be detected in the kurtosis of the symmetric \mexhat\ wavelet coefficients on scale $\scale_9=450\arcmin$ and in the kurtosis of the elliptical \mexhat\ wavelet coefficients on scale $\scale_{12}=600\arcmin$.  These new candidate detections are investigated further in the next section.
Interestingly, the skewness detections that we previously made are not mitigated when making the Bianchi correction -- the highly significant detection of non-Gaussianity previously made with the \morlet\ wavelet remains.

It is also interesting to note that the co-added \wmap\ and \ilc\ maps both exhibit similar statistics, suggesting it is appropriate to use the Bianchi template fitted to the \ilc\ map to correct the co-added \wmap\ map.


\newlength{\statplotwidth}
\setlength{\statplotwidth}{55mm}

\begin{figure*}
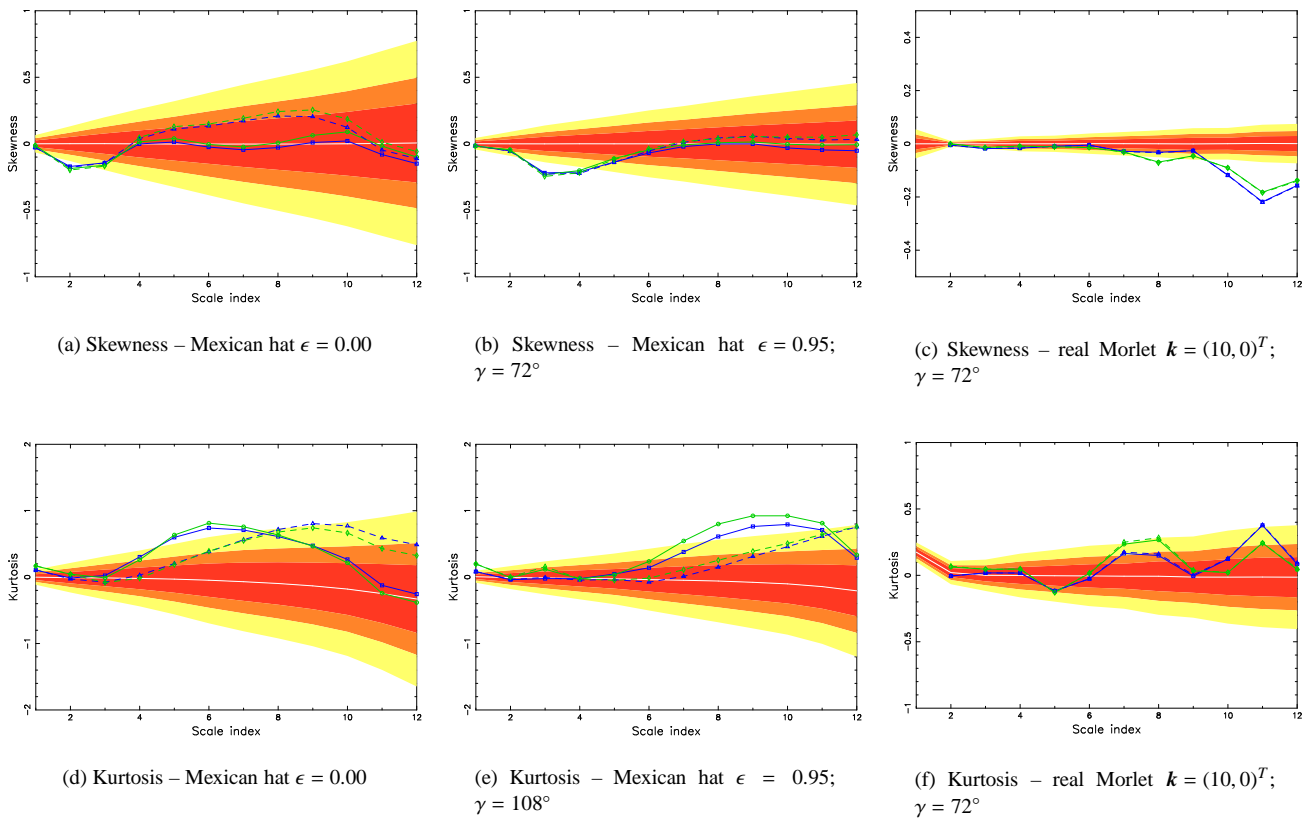

\begin{minipage}{\textwidth}
\centering
\mbox{
\subfigure[Skewness -- \mexhat\ $\eccen=0.00$]
  {\includegraphics[trim=0mm 0mm -3mm 0mm,clip,angle=-90,width=\statplotwidth]{figures/skewness_mexhat000_ig01}} \quad
\subfigure[Skewness -- \mexhat\ \mbox{$\eccen=0.95$}; \mbox{$\eulerc=72^\circ$}]
  {\includegraphics[trim=0mm 0mm -3mm  0mm,clip,angle=-90,width=\statplotwidth]{figures/skewness_mexhat095_ig02}} \quad
\subfigure[Skewness -- \morlet\ \mbox{$\bmath{k}=\left( 10, 0 \right)^{T}$}; \mbox{$\eulerc=72^\circ$}]
  {\includegraphics[trim=0mm 0mm -3mm 0mm,clip,angle=-90,width=\statplotwidth]{figures/skewness_morlet_ig02}}
}
\mbox{
\subfigure[Kurtosis -- \mexhat\ $\eccen=0.00$]
  {\includegraphics[trim=0mm 0mm -3mm 0mm,clip,angle=-90,width=\statplotwidth]{figures/kurtosis_mexhat000_ig01}} \quad
\subfigure[Kurtosis -- \mexhat\ $\eccen=0.95$; \mbox{$\eulerc=108^\circ$}]
  {\includegraphics[trim=0mm 0mm -3mm 0mm,clip,angle=-90,width=\statplotwidth]{figures/kurtosis_mexhat095_ig05}} \quad
\subfigure[Kurtosis -- \morlet\ \mbox{$\bmath{k}=\left( 10, 0 \right)^{T}$}; \mbox{$\eulerc=72^\circ$}]
  {\includegraphics[trim=0mm 0mm -3mm 0mm,clip,angle=-90,width=\statplotwidth]{figures/kurtosis_morlet_ig02}}
}
\caption{Spherical wavelet coefficient statistics for each wavelet and map.  Confidence regions
  obtained from \ngsim\ Monte Carlo simulations are shown for 68\% (red/dark-grey), 95\%
  (orange/grey) and 99\% (yellow/light-grey) levels, as is the mean (solid white
  line). 
  Statistics corresponding to the following maps are plotted: 
  \wmap\ combined map (solid, blue, squares);
  ILC map (solid, green, circles); 
  Bianchi corrected \wmap\ combined map (dashed, blue, triangles); 
  Bianchi corrected ILC map (dashed, green, diamonds).
  Only the orientations corresponding to the most significant deviations
  from Gaussianity are shown for the \mexhat\ $\eccen=0.95$ and
  \morlet\ wavelet cases.}
\label{fig:stat_plot}
\end{minipage}
\end{figure*}

\subsection{Statistical significance of detections}
\label{sec:stat_sig}

We examine the statistical significance of deviations from Gaussianity in more detail.  Our first approach is to examine the distribution of the statistics that show the most significant deviation from Gaussianity in the uncorrected maps, in order to associate significance levels with the detections.
Our second approach is to perform $\chi^2$ tests on the statistics computed with each type of spherical wavelet.  This approach considers all statistics in aggregate, and infers a significance level for deviations from Gaussianity in the entire set of test statistics.

Histograms constructed from the Monte Carlo simulations for those test statistics corresponding to the most significant deviations from Gaussianity are shown in \fig{\ref{fig:hist}}.
The measured statistic of each map considered is also shown on the plots, with the number of standard deviations these observations deviate from the mean.
Notice that the deviation from the mean of the kurtosis statistics is considerably reduced in the Bianchi corrected maps, whereas the deviation for the skewness statistics is not significantly affected.

\newif\ifnotext
\notextfalse

\ifnotext

  \begin{figure*}
  \begin{minipage}{\textwidth}
  \centering
  \mbox{
  \subfigure[Skewness -- \mexhat\ $\eccen=0.00$; \mbox{$\scale_2=100\arcmin$}]
    {\includegraphics[trim=0mm 0mm -3mm 0mm,clip,angle=-90,width=\statplotwidth]{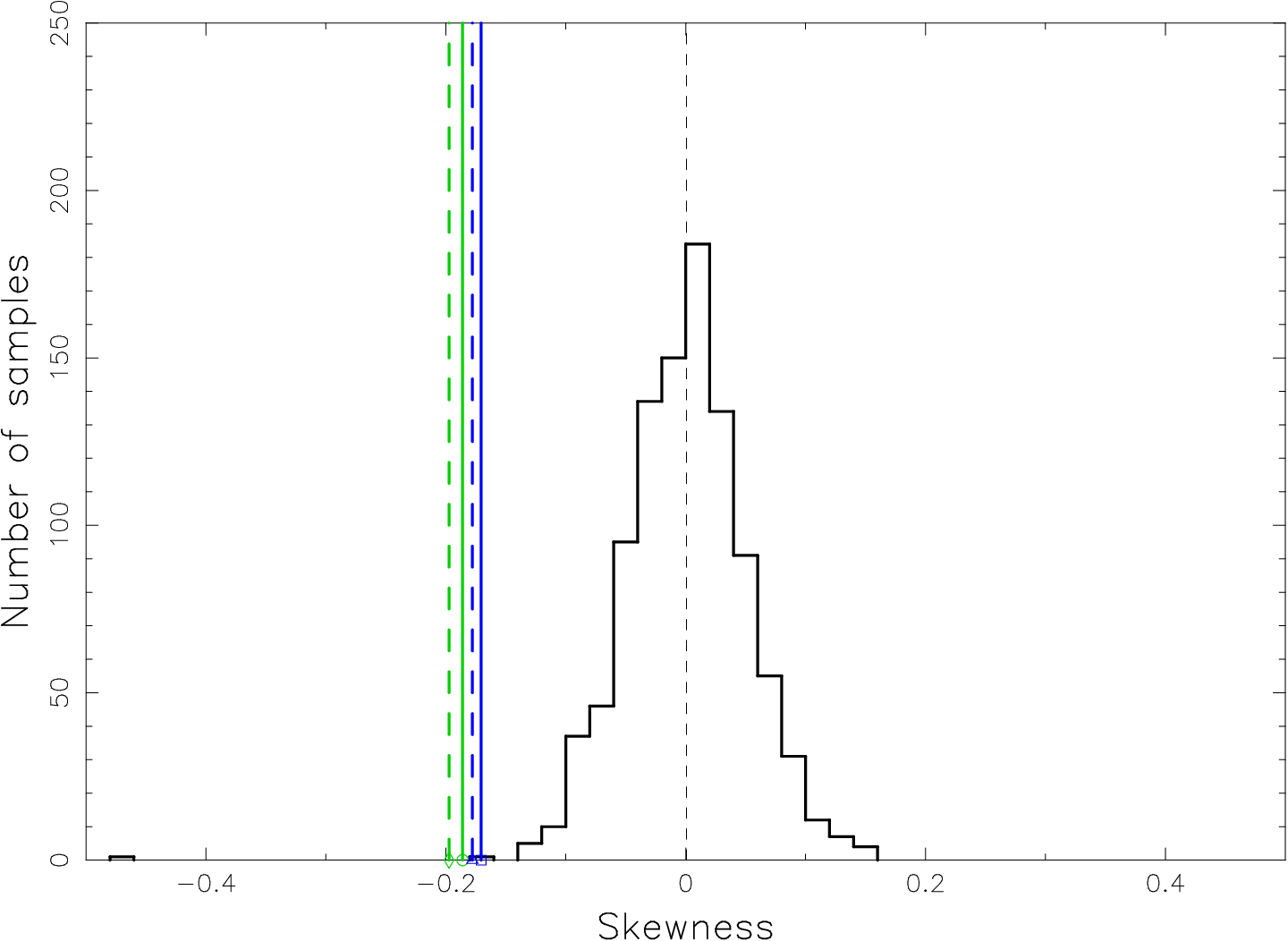}} \quad
  \subfigure[Skewness -- \mexhat\ \mbox{$\eccen=0.95$}; $\scale_3=150\arcmin$; $\eulerc=72^\circ$]
    {\includegraphics[trim=0mm 0mm -3mm 0mm,clip,angle=-90,width=\statplotwidth]{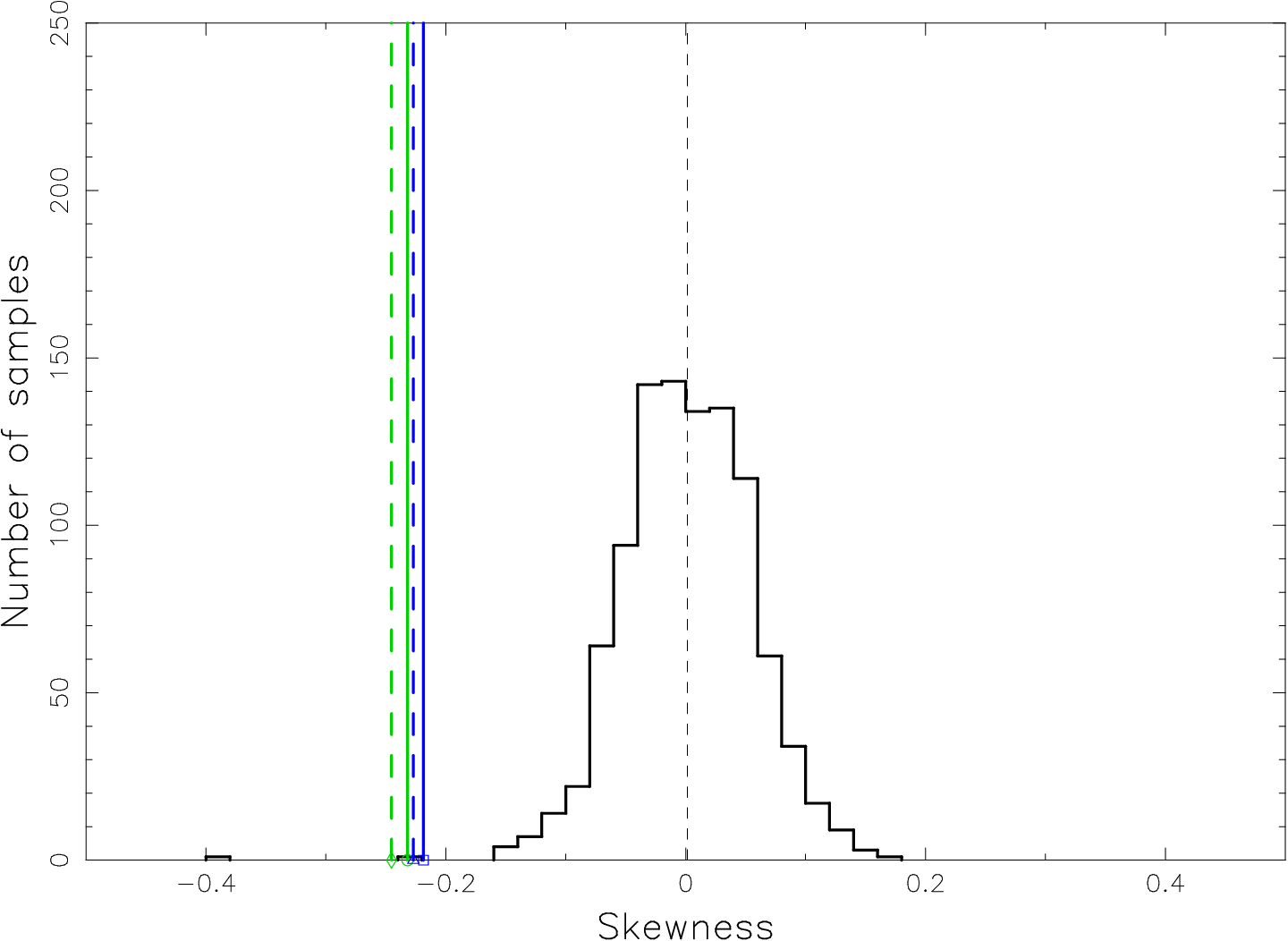}} \quad
  \subfigure[Skewness -- \morlet\ \mbox{$\bmath{k}=\left( 10, 0 \right)^{T}$}; $\scale_{11}=550\arcmin$; $\eulerc=72^\circ$]
    {\includegraphics[trim=0mm 0mm -3mm 0mm,clip,angle=-90,width=\statplotwidth]{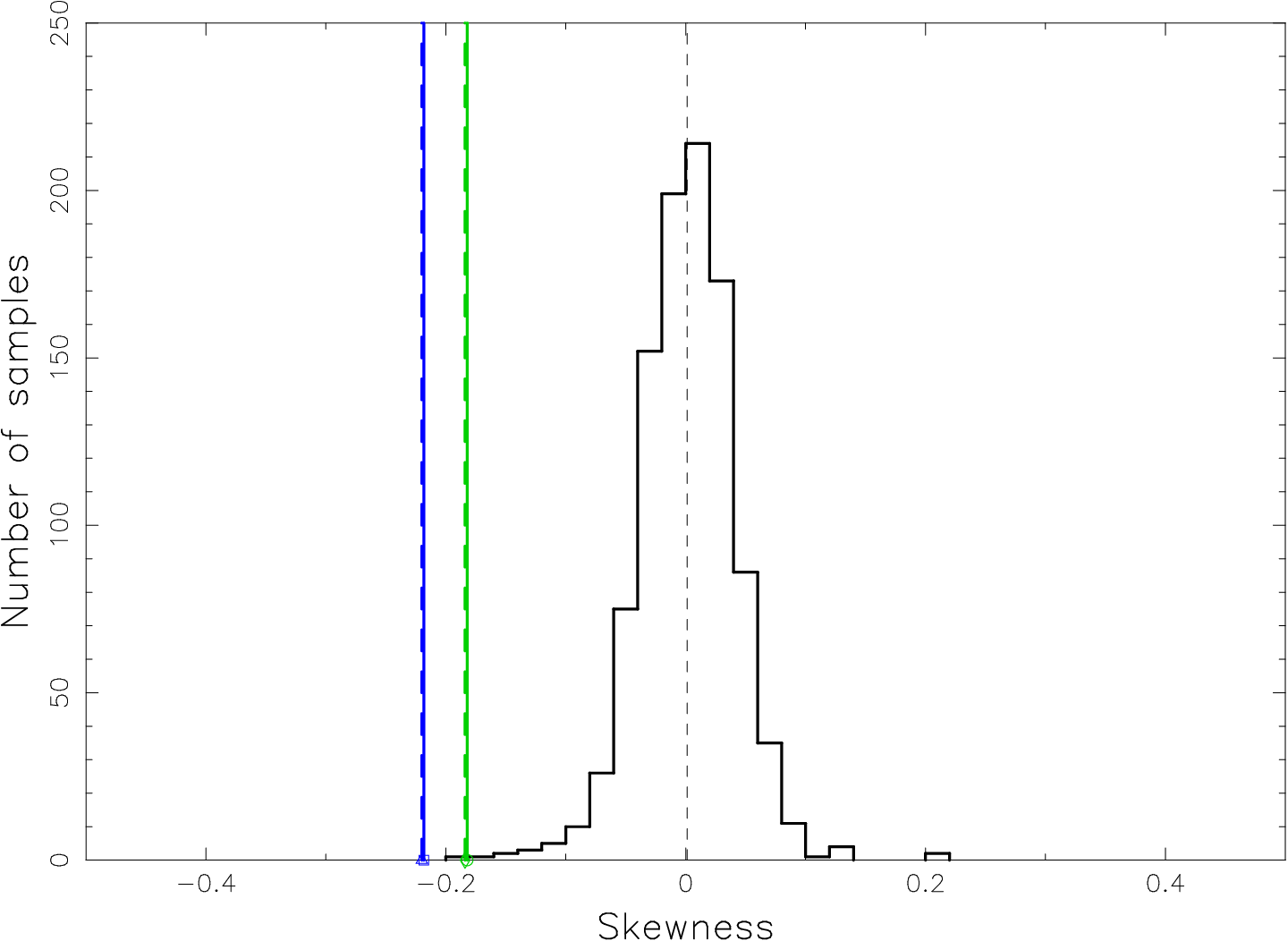}}
  }
  \mbox{
  \subfigure[Kurtosis -- \mexhat\ $\eccen=0.00$, \mbox{$\scale_6=300\arcmin$}]
    {\includegraphics[trim=0mm 0mm -3mm 0mm,clip,angle=-90,width=\statplotwidth]{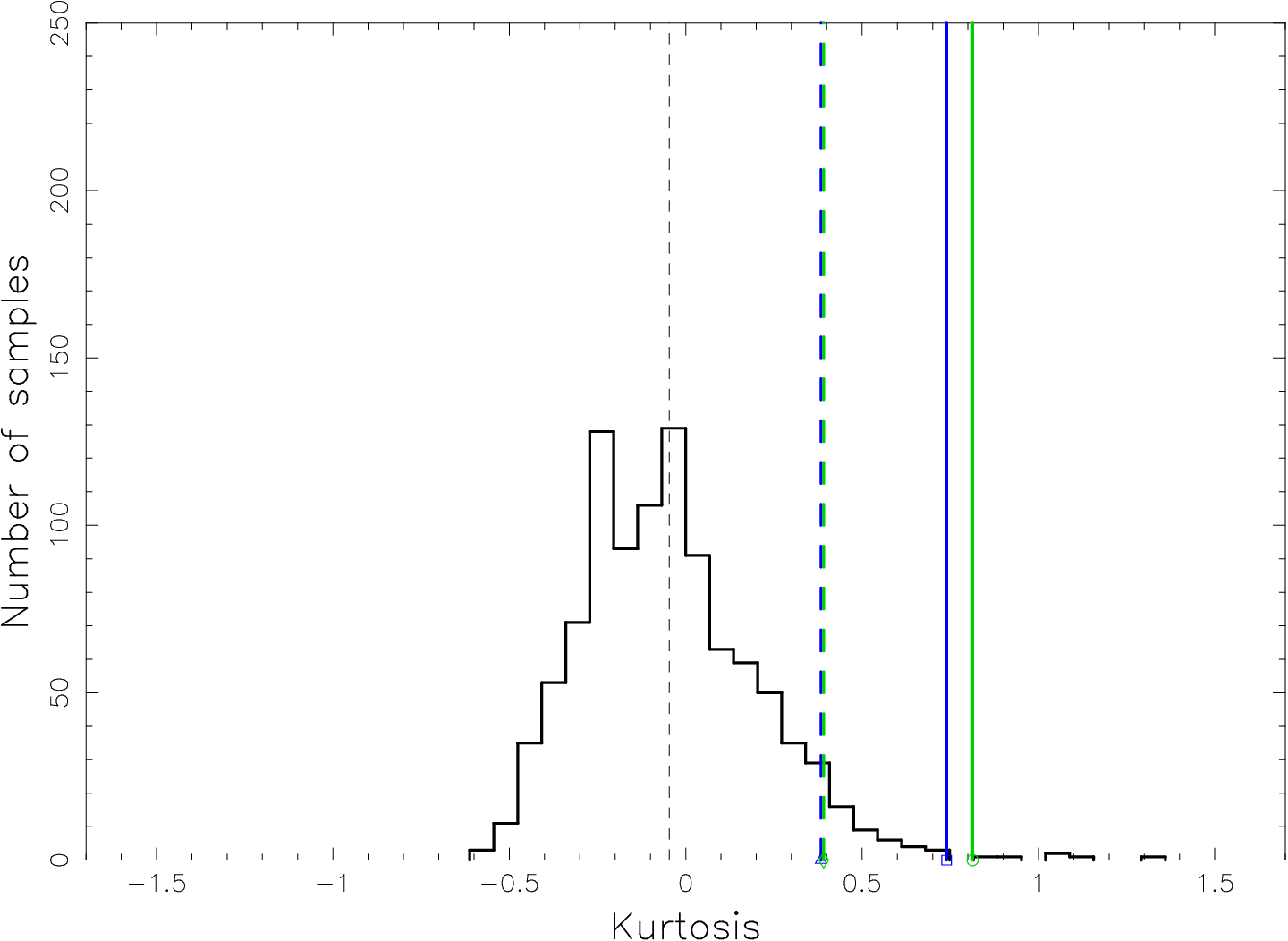}} \quad
  \subfigure[Kurtosis -- \mexhat\ $\eccen=0.95$; \mbox{$\scale_{10}=500\arcmin$}; $\eulerc=108^\circ$]
    {\includegraphics[trim=0mm 0mm -3mm 0mm,clip,angle=-90,width=\statplotwidth]{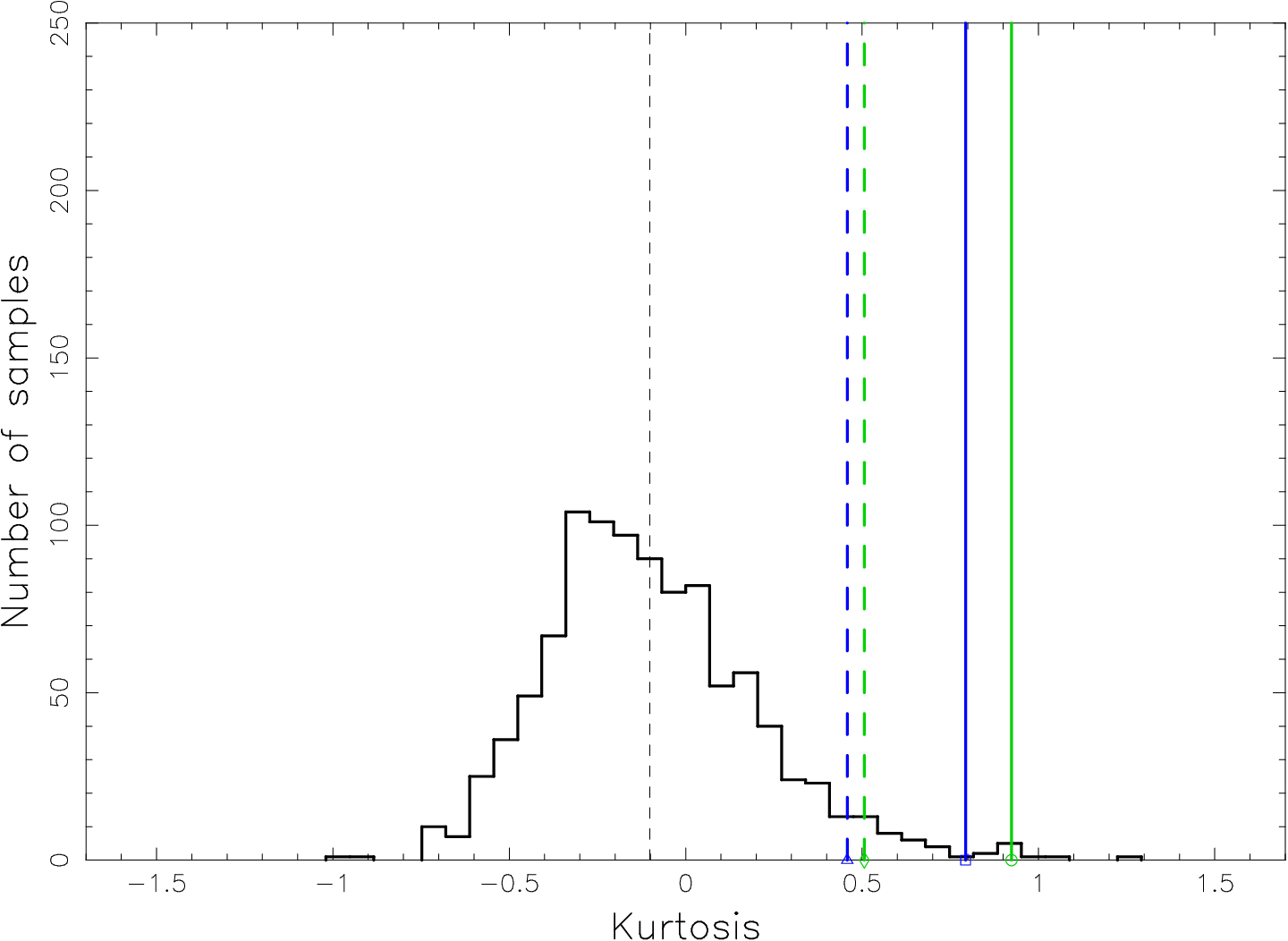}} \quad
  \subfigure[Kurtosis -- \morlet\ \mbox{$\bmath{k}=\left( 10, 0 \right)^{T}$}; \mbox{$\scale_{11}=550\arcmin$}; $\eulerc=72^\circ$]
    {\includegraphics[trim=0mm 0mm -3mm 0mm,clip,angle=-90,width=\statplotwidth]{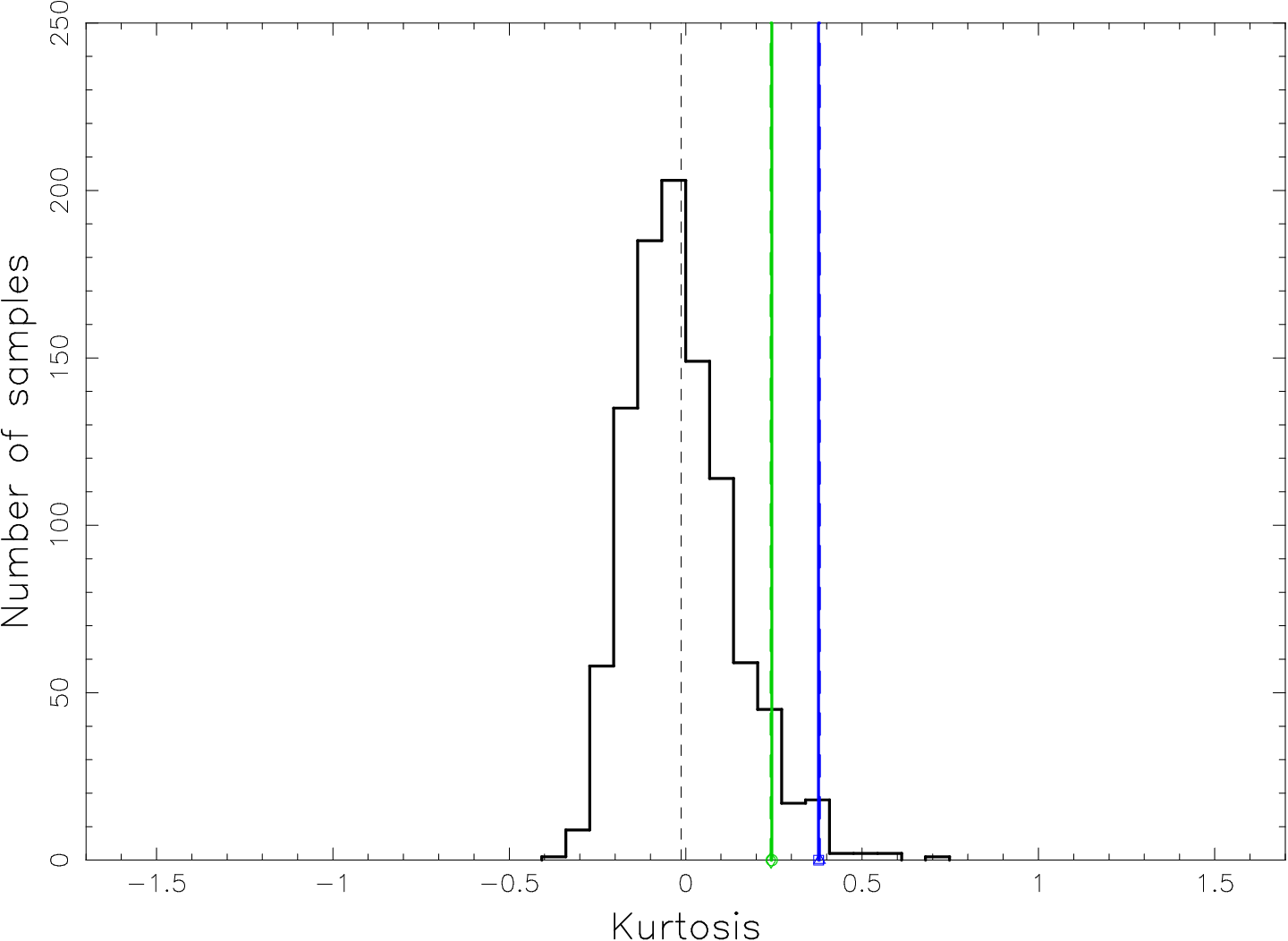}}
  }
  \caption{Histograms of spherical wavelet coefficient statistic
    obtained from \ngsim\ Monte Carlo simulations. The mean is shown by
    the thin dashed black vertical line.
    Observed statistic corresponding to the following maps are plotted:
    \wmap\ combined map (solid, blue, square);
    ILC map (solid, green, circle);
    Bianchi corrected \wmap\ combined map (dashed, blue, triangle);
    Bianchi corrected ILC map (dashed, green, diamond).
    Only those scales and orientations corresponding to the most
    significant deviations from Gaussianity are shown for each wavelet.}
  \label{fig:hist}
  \end{minipage}
  \end{figure*}

\else

  \begin{figure*}
  \begin{minipage}{\textwidth}
  \centering
  \mbox{
  \subfigure[Skewness -- \mexhat\ $\eccen=0.00$; \mbox{$\scale_2=100\arcmin$}]
    {\includegraphics[trim=0mm 0mm -3mm 0mm,clip,angle=-90,width=\statplotwidth]{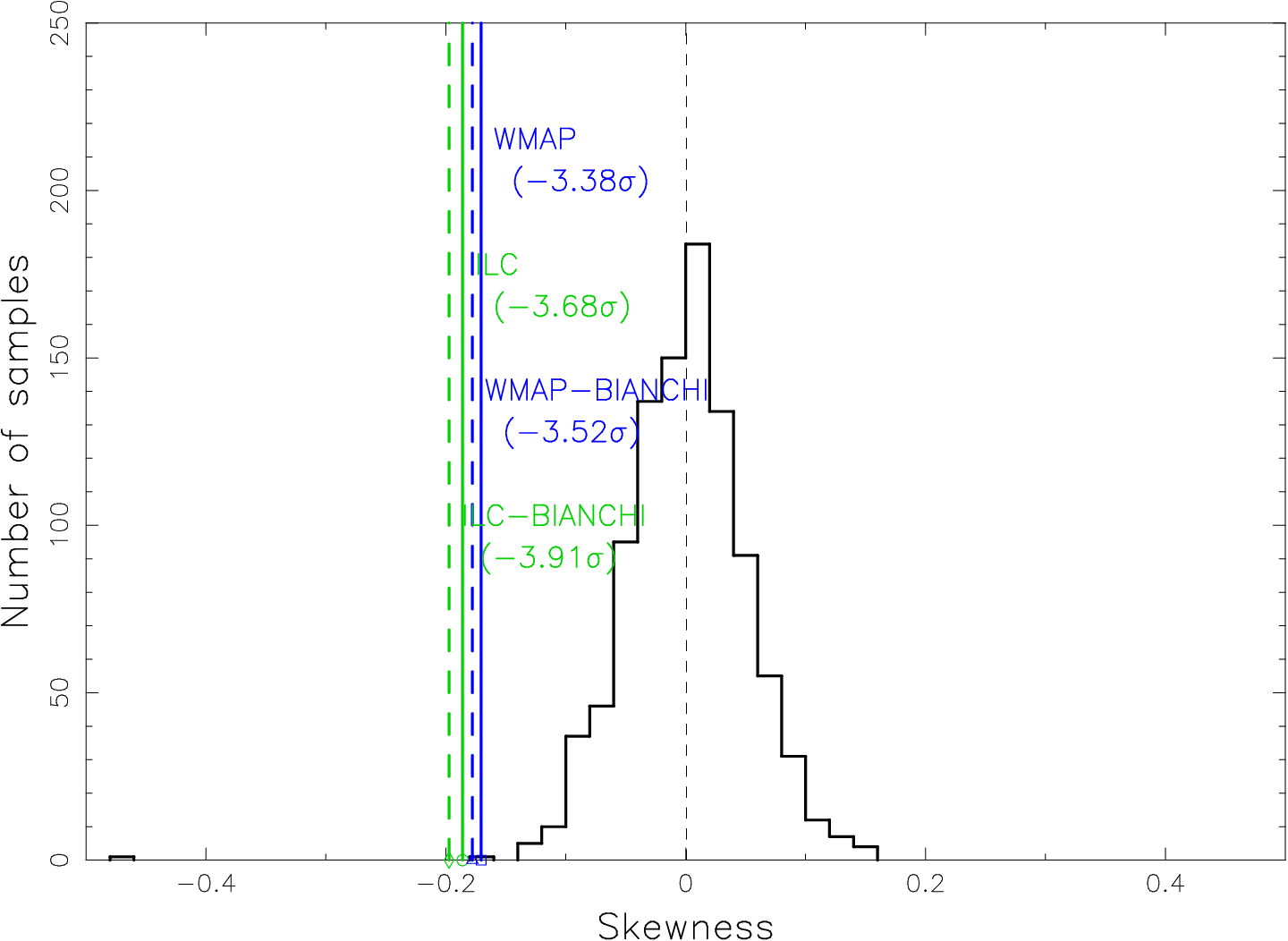}} \quad
  \subfigure[Skewness -- \mexhat\ \mbox{$\eccen=0.95$}; $\scale_3=150\arcmin$; $\eulerc=72^\circ$]
    {\includegraphics[trim=0mm 0mm -3mm 0mm,clip,angle=-90,width=\statplotwidth]{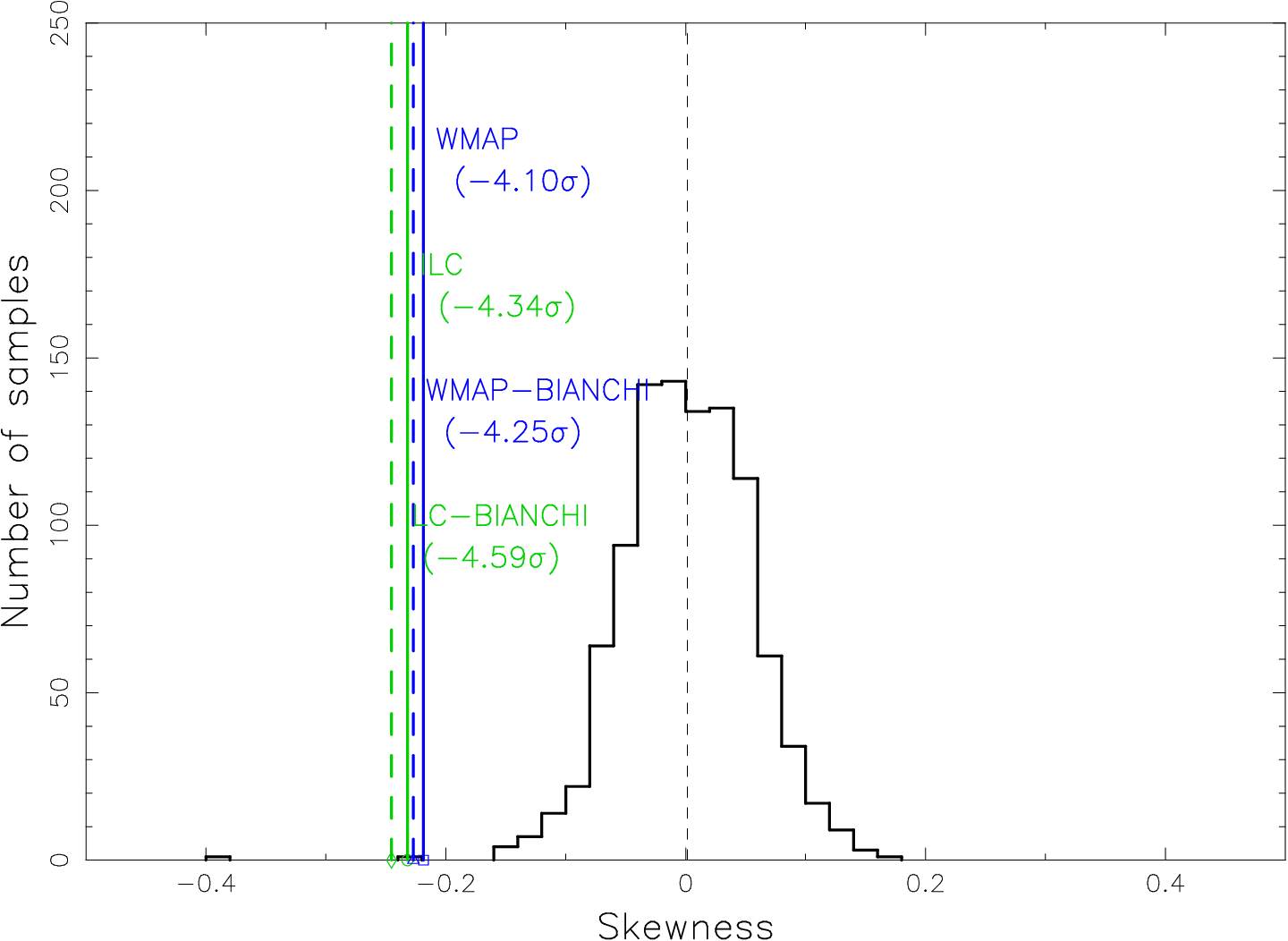}} \quad
  \subfigure[Skewness -- \morlet\ \mbox{$\bmath{k}=\left( 10, 0 \right)^{T}$}; $\scale_{11}=550\arcmin$; $\eulerc=72^\circ$]
    {\includegraphics[trim=0mm 0mm -3mm 0mm,clip,angle=-90,width=\statplotwidth]{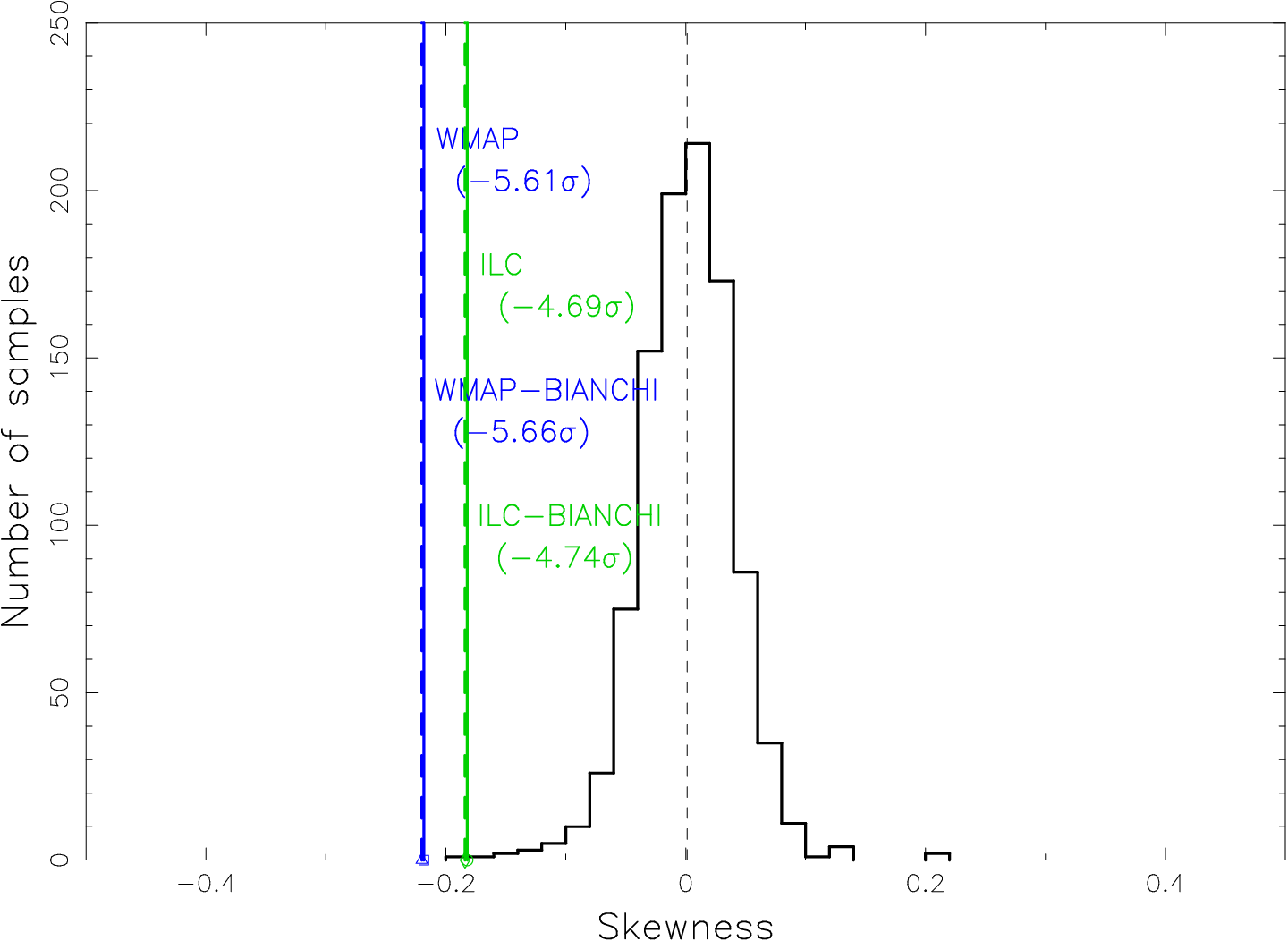}}
  }
  \mbox{
  \subfigure[Kurtosis -- \mexhat\ $\eccen=0.00$, \mbox{$\scale_6=300\arcmin$}]
    {\includegraphics[trim=0mm 0mm -3mm 0mm,clip,angle=-90,width=\statplotwidth]{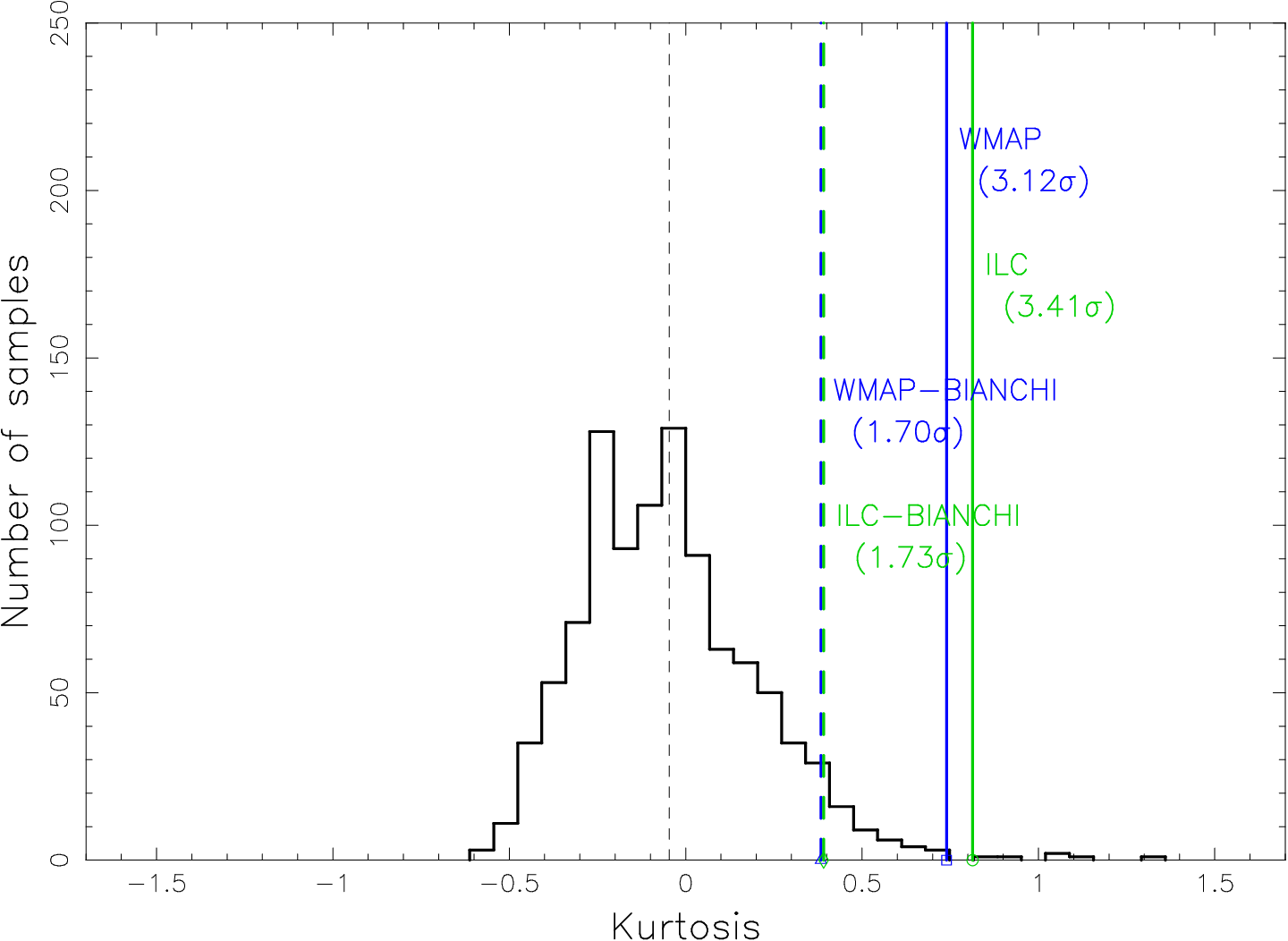}} \quad
  \subfigure[Kurtosis -- \mexhat\ $\eccen=0.95$; \mbox{$\scale_{10}=500\arcmin$}; $\eulerc=108^\circ$]
    {\includegraphics[trim=0mm 0mm -3mm 0mm,clip,angle=-90,width=\statplotwidth]{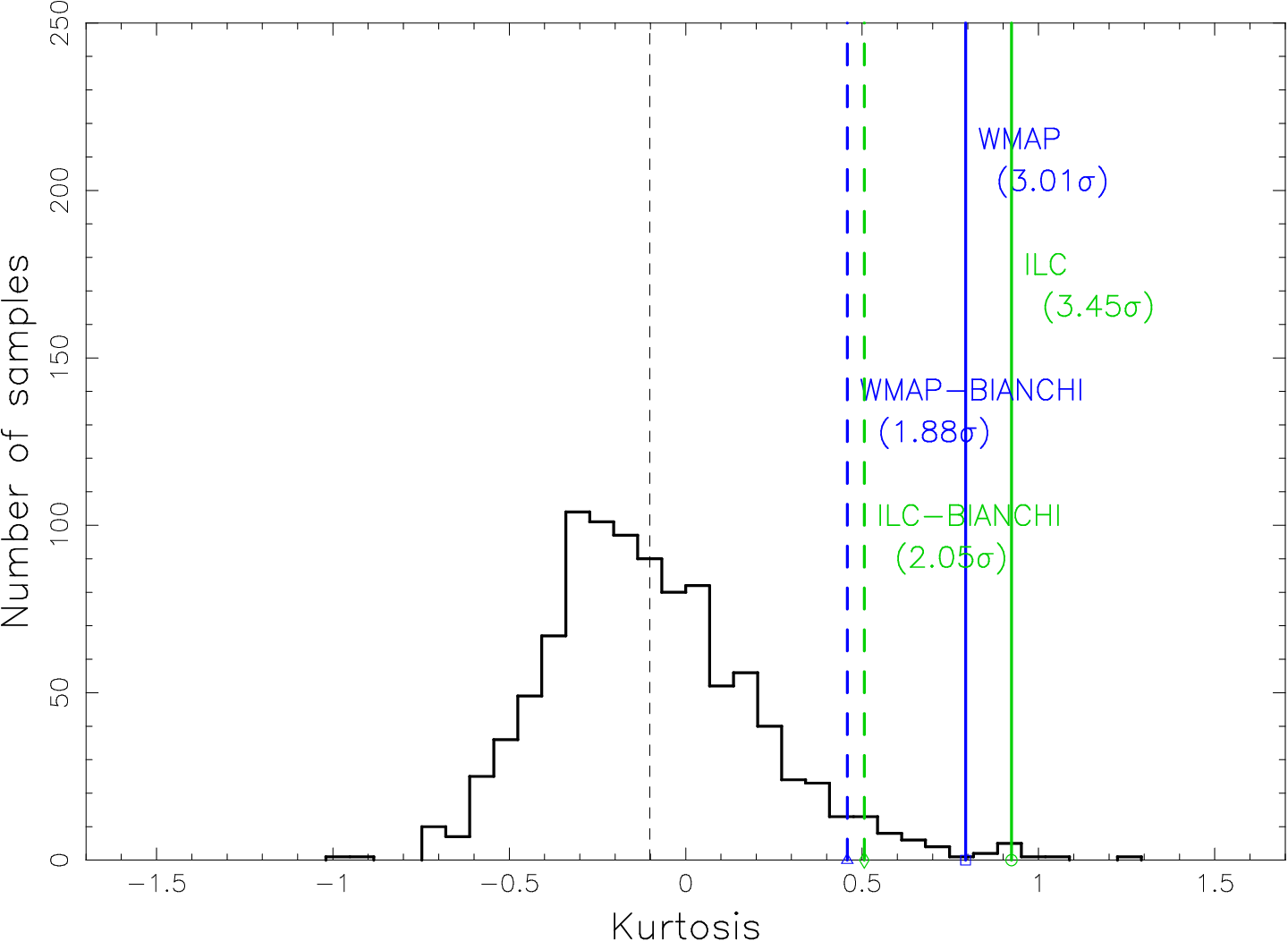}} \quad
  \subfigure[Kurtosis -- \morlet\ \mbox{$\bmath{k}=\left( 10, 0 \right)^{T}$}; \mbox{$\scale_{11}=550\arcmin$}; $\eulerc=72^\circ$]
    {\includegraphics[trim=0mm 0mm -3mm 0mm,clip,angle=-90,width=\statplotwidth]{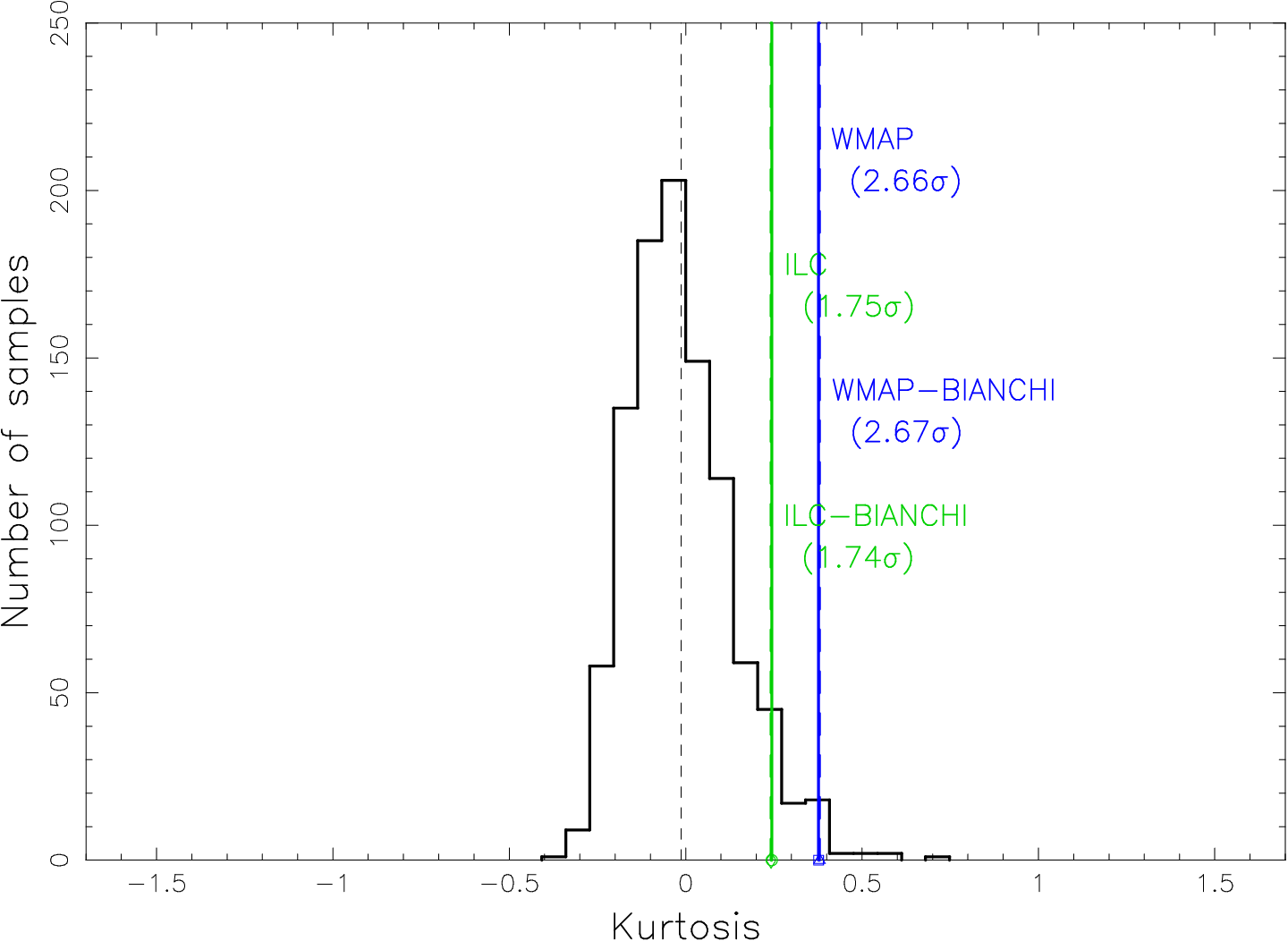}}
  }
  \caption{Histograms of spherical wavelet coefficient statistic
    obtained from \ngsim\ Monte Carlo simulations. The mean is shown by
    the thin dashed black vertical line.  
    Observed statistics corresponding to the following maps are also plotted:
    \wmap\ combined map (solid, blue, square);
    ILC map (solid, green, circle); 
    Bianchi corrected \wmap\ combined map (dashed, blue, triangle); 
    Bianchi corrected ILC map (dashed, green, diamond).
    The number of standard deviations these observations
    deviate from the mean is also displayed on each plot.
    Only those scales and orientations corresponding to the most
    significant deviations from Gaussianity are shown for each wavelet.}
  \label{fig:hist}
  \end{minipage}
  \end{figure*}

\fi

Next we construct significance measures for each of the most significant detections of non-Gaussianity.
For each wavelet, we determine the probability that \emph{any} single statistic (either skewness or kurtosis) in the Monte Carlo simulations deviates by an equivalent or greater amount than the test statistic under examination.
If any skewness or kurtosis statistic%
\footnote{
Although we recognise the distinction between skewness and kurtosis, there is no reason to partition the set of test statistics into skewness and kurtosis subsets.  The full set of test statistics must be considered.}
calculated from the simulated Gaussian map -- on any scale or orientation -- deviates more than the maximum deviation observed in the data map for that wavelet, then the map is flagged as exhibiting a more significant deviation. 
This technique is an extremely conservative means of constructing significance levels for the observed test statistics.
We use the number of standard deviations to characterise the deviation of the detections, rather that the exact probability given by the simulations, since for many of the statistics we consider no simulations exhibit as great a deviation on the particular scale and orientation.  Using the number of standard deviations is therefore a more robust approach and is consistent with our previous work \citep{mcewen:2005a}.
Significance levels corresponding to the detections considered in \fig{\ref{fig:hist}} are calculated and displayed in \tbl{\ref{tbl:num_deviations}}.
For clarity, we show only those values from the co-added \wmap\ map, although the ILC map exhibits similar results.
These results confirm our inferences from direct observation of the statistics relative to the confidence levels and histograms shown in \fig{\ref{fig:stat_plot}} and \fig{\ref{fig:hist}} respectively: the original kurtosis detections of non-Gaussianity are eliminated, while the original skewness detections remain.
We also determine the significance of the new candidate detections of non-Gaussianity observed in the kurtosis of the \mexhat\ wavelet coefficients in the Bianchi corrected data.  Of the 1000 simualtions, 115 contain a statistic that exhibits a greater deviation that the symmetric \mexhat\ wavelet kurtosis on scale $\scale_9$ of the Bianchi corrected data, hence this detection may only be made at the 88.5\% significance level.  448 of the simulations contain a statistic that exhibits a greater deviation that the elliptical \mexhat\ wavelet kurtosis on scale $\scale_{12}$ of the Bianchi corrected data, hence this candidate detection may only be made at the 55.2\% significance level.  Thus we conclude that no highly significant detection of non-Gaussianity can be made on any scale or orientation from the analysis of the kurtosis of spherical wavelet coefficients in the Bianchi corrected data, however the detections made previously in the skewness of spherical wavelet coefficients remain essentially unaltered.

\begin{table}
  \begin{center}
  \caption{Deviation and significance levels of spherical wavelet
    coefficient statistics calculated from the \wmap\ and Bianchi
    corrected \wmap\ maps (similar
    results are obtained using the ILC map).
    Standard deviations and significant levels 
    are calculated from \ngsim\ Monte Carlo simulations.
    The table variables are defined as follows: the number of standard
    deviations the observation deviates from the mean is given by \nstd;
    the number of simulated Gaussian maps that exhibit an equivalent or greater deviation
    in \emph{any} test statistics calculated using the given wavelet is
    given by \ndev; the corresponding significance level of the
    non-Gaussianity detection is given by \conflevel.
    Only those scales and orientations corresponding to the most
    significant deviations from Gaussianity are listed for each wavelet.}
  \label{tbl:num_deviations}
  \subfigure[\Mexhat\ $\eccen=0.00$]
  {
  \begin{tabular}{lcccc} \hline
  & \multicolumn{2}{c}{Skewness} & \multicolumn{2}{c}{Kurtosis} \\
  & \multicolumn{2}{c}{($\scale_2=100\arcmin$)} & \multicolumn{2}{c}{($\scale_6=300\arcmin$)} \\
  & \wmap & \wmapbianchi & \wmap & \wmapbianchi \\ \hline
  \nstd & \nstdmexskewsgn & \nstdmexskewsgnbian & \nstdmexkurtsgn & \nstdmexkurtsgnbian \\
  \ndev & \nstatmexskew\ maps & \nstatmexskewbian\ maps & \nstatmexkurt\ maps & \nstatmexkurtbian\ maps  \\
  \conflevel & \clmexskew\% & \clmexskewbian\% & \clmexkurt\% & \clmexkurtbian\% \\ \hline
  \end{tabular}
  }
  \subfigure[\Mexhat\ $\eccen=0.95$]
  {
  \begin{tabular}{lcccc} \hline
  & \multicolumn{2}{c}{Skewness} & \multicolumn{2}{c}{Kurtosis} \\
  & \multicolumn{2}{c}{($\scale_3=150\arcmin$; $\eulerc=72^\circ$)} & \multicolumn{2}{c}{($\scale_{10}=500\arcmin$; $\eulerc=108^\circ$)}\\
  & \wmap & \wmapbianchi & \wmap & \wmapbianchi \\ \hline
  \nstd & \nstdmexepskewsgn & \nstdmexepskewsgnbian & \nstdmexepkurtsgn & \nstdmexepkurtsgnbian \\
  \ndev & \nstatmexepskew\ maps & \nstatmexepskewbian\ maps & \nstatmexepkurt\ maps & \nstatmexepkurtbian\ maps \\
  \conflevel & \clmexepskew\% & \clmexepskewbian\% & \clmexepkurt\% & \clmexepkurtbian\% \\ \hline
  \end{tabular}
  }
  \subfigure[\Morlet\ $\bmath{k}=\left( 10, 0 \right)^{T}$]
  {
  \begin{tabular}{lcccc} \hline
  & \multicolumn{2}{c}{Skewness} & \multicolumn{2}{c}{Kurtosis} \\
  & \multicolumn{2}{c}{($\scale_{11}=550\arcmin$; $\eulerc=72^\circ$)} & \multicolumn{2}{c}{($\scale_{11}=550\arcmin$; $\eulerc=72^\circ$)} \\
  & \wmap & \wmapbianchi & \wmap & \wmapbianchi \\ \hline
  \nstd & \nstdmorskewsgn & \nstdmorskewsgnbian & \nstdmorkurtsgn & \nstdmorkurtsgnbian \\
  \ndev & \nstatmorskew\ maps & \nstatmorskewbian\ maps & \nstatmorkurt\ maps & \nstatmorkurtbian\ maps \\
  \conflevel & \clmorskew\% & \clmorskewbian\% & \clmorkurt\% & \clmorkurtbian\% \\ \hline
  \end{tabular}
  }
  \end{center}
\end{table}

Finally, we perform $\chi^2$ tests to probe the significance of deviations from Gaussianity in the aggregate set of test statistics.  
These tests inherently take all statistics on all scales and orientations into account.
The results of these tests are summarised in \fig{\ref{fig:chisqd}}.
The overall significance of the detection of non-Gaussianity is reduced for the \mexhat\ wavelets, although this reduction is not as marked as that illustrated in \tbl{\ref{tbl:num_deviations}} since both skewness and kurtosis statistics are considered when computing the $\chi^2$, and it is only the kurtosis detection that is eliminated.  For example, when an equivalent $\chi^2$ test is performed using only the kurtosis statistics the significance of the detection made with the symmetric \mexhat\ wavelet drops from $99.9\%$ to $95\%$ (note that this is still considerably higher than the level found with the previous test, illustrating just how conservative the previous method is).
The significance of the detection made with the \morlet\ wavelet is not
affected by correcting for the Bianchi template.
This is expected since the detection was made only in the skewness of the \morlet\ wavelet coefficients and not the kurtosis.

We quote the overall significance of our detections of non-Gaussianity at the level calculated by the first approach, since this is the more conservative of the two tests.

\newlength{\chiplotwidth}
\setlength{\chiplotwidth}{72mm}

\begin{figure}
\centering
\subfigure[\Mexhat\ $\eccen=0.00$]{\includegraphics[trim=0mm 0mm -3mm 0mm,clip,angle=-90,width=\chiplotwidth]{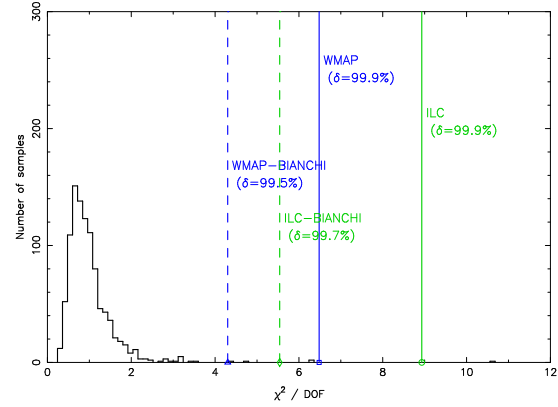}}
\subfigure[\Mexhat\ $\eccen=0.95$]{\includegraphics[trim=0mm 0mm -3mm 0mm,clip,angle=-90,width=\chiplotwidth]{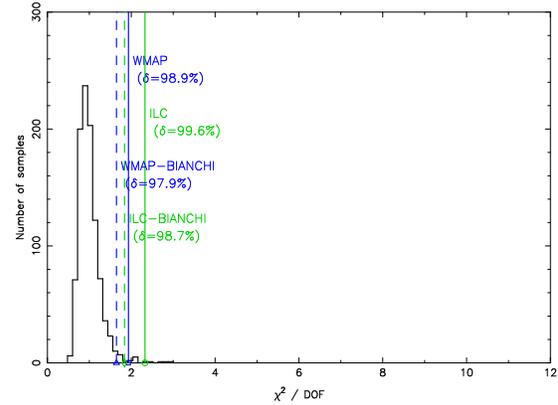}}
\subfigure[\Morlet\ $\bmath{k}=\left( 10, 0 \right)^{T}$]{\includegraphics[trim=0mm 0mm -3mm 0mm,clip,angle=-90,width=\chiplotwidth]{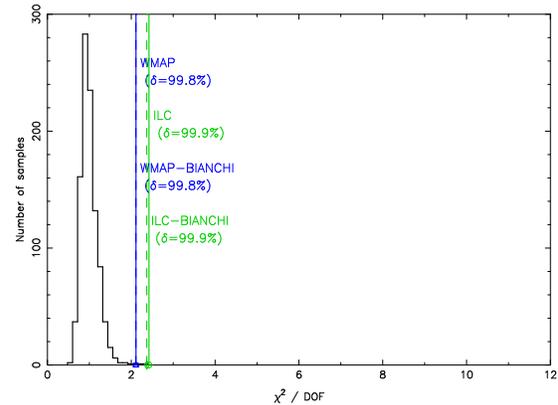}}
\caption{Histograms of normalised $\chi^2$ test
  statistics obtained from \ngsim\ Monte Carlo simulations.
  Normalised $\chi^2$ values corresponding to the following maps are
  also plotted:
  \wmap\ combined map (solid, blue, square);
  ILC map (solid, green, circle);
  Bianchi corrected \wmap\ combined map (dashed, blue, triangle);
  Bianchi corrected ILC map (dashed, green, diamond).
  The significance level of each detection made using $\chi^2$ values
  is also quoted ($\delta$).}
\label{fig:chisqd}
\end{figure}


\subsection{Localised deviations from Gaussianity}

The spatial localisation inherent in the wavelet analysis allows one to localise most likely sources of non-Gaussianity on the sky.  We examine spherical wavelet coefficients maps thresholded so that those coefficients below $3\sigma$ (in absolute value) are set to zero.  The remaining coefficients show likely regions that contribute to deviations from Gaussianity in the map.

The localised regions of the skewness-flagged maps for each wavelet are almost identical for all of the original and Bianchi corrected co-added \wmap\ and \ilc\ maps.
This is expected as it has been shown that the Bianchi correction does not remove the skewness detection.
All of these thresholded coefficient maps are almost identical to those shown in \fig{9~(a,c,d)} of \citet{mcewen:2005a}, hence they are not shown here.

The localised regions detected in the kurtosis-flagged maps for the \mexhat\ wavelets are shown in \fig{\ref{fig:coeff}} (the real Morlet wavelet did not flag a significant kurtosis detection of non-Gaussianity).
The thresholded coefficient maps for the original and Bianchi corrected data are reasonably similar, however the size and magnitude of the cold spot at Galactic coordinates \spotloc\ is significantly reduced in the Bianchi corrected maps.  \citet{cruz:2005} claim that it is this cold spot that is responsible for the kurtosis detection of non-Gaussianity.  This may explain why the kurtosis detection of non-Gaussianity is eliminated in the Bianchi corrected maps.

Although the new cadidate detections of non-Gaussianity observed in the kurtosis of the \mexhat\ wavelet coefficients of the Bianchi corrected data were shown in \sectn{\ref{sec:stat_sig}} not to be particularly significant, we nevertheless construct the localised maps corresponding to these candidate detections.  The regions localised in these maps show no additional structure than that shown in \fig{\ref{fig:coeff}}.  The only significant difference between the localised regions is that the cold spot at \spotloc\ is absent.

\newlength{\coeffplotwidth}
\setlength{\coeffplotwidth}{52mm}

\begin{figure*}
\begin{minipage}{\textwidth}
\centering
\mbox{
  \subfigure[Co-added \wmap\ map \mexhat\ wavelet coefficients ($\eccen=0.00$; \mbox{$\scale_6=300\arcmin$)}]
    { \hspace{5mm}
      \begin{minipage}[t]{55mm}
        \vspace{0pt}
        \includegraphics[width=\coeffplotwidth]{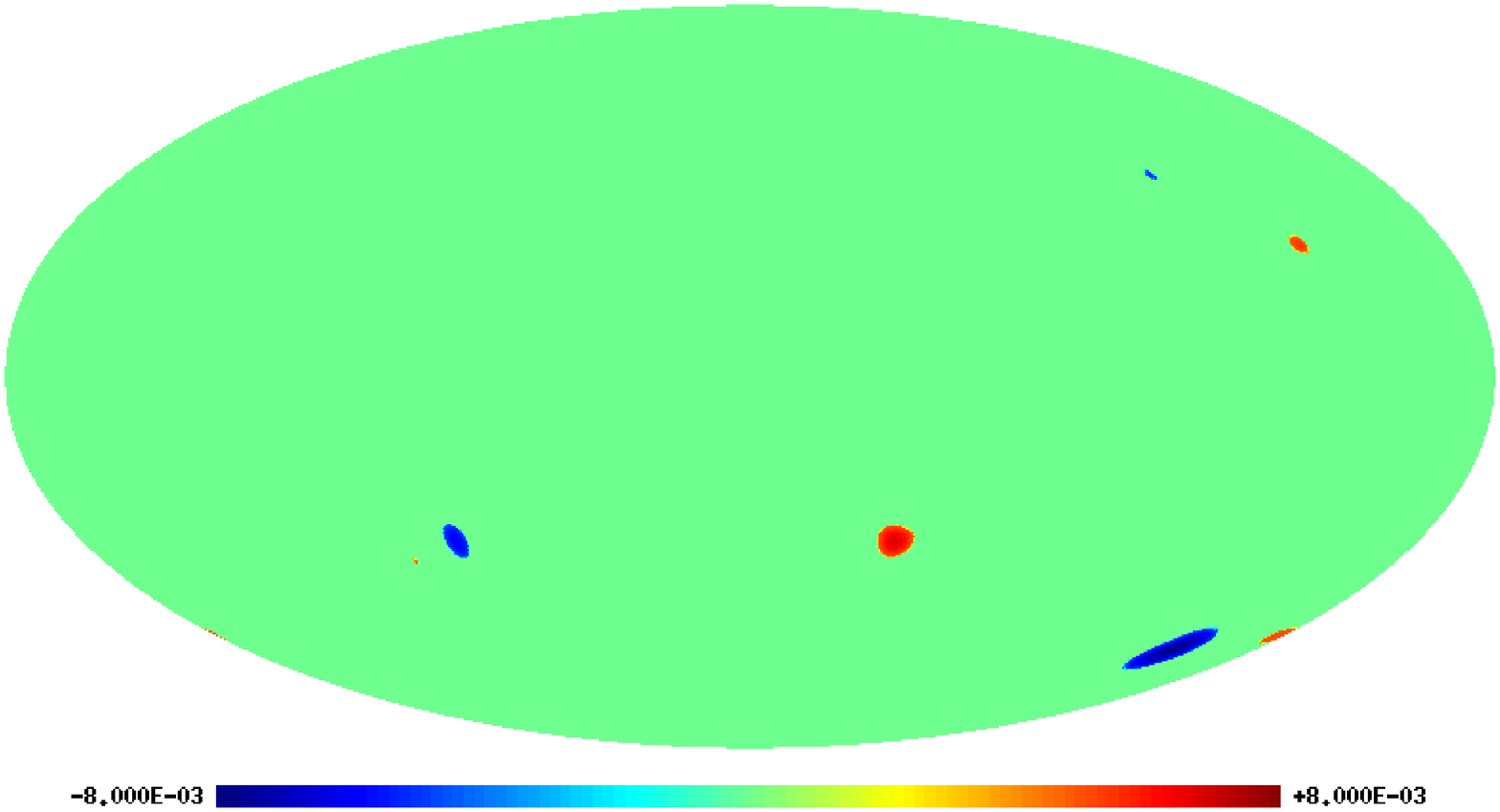}
      \end{minipage}\hspace{-3mm}
      \begin{minipage}[t]{25mm}
        \vspace{0pt}
        \frame{\includegraphics[bb= 570 50 700 120,width=20mm,clip]{figures/wmap_wcoeff_mexhat000_thres_ia06_ig01}}
      \end{minipage}
    }
  \hspace{5mm}
  \subfigure[Bianchi corrected co-added \wmap\ map \mexhat\ wavelet coefficients ($\eccen=0.00$; $\scale_6=300\arcmin$)]
    { \hspace{5mm}
      \begin{minipage}[t]{55mm}
        \vspace{0pt}
        \includegraphics[width=\coeffplotwidth]{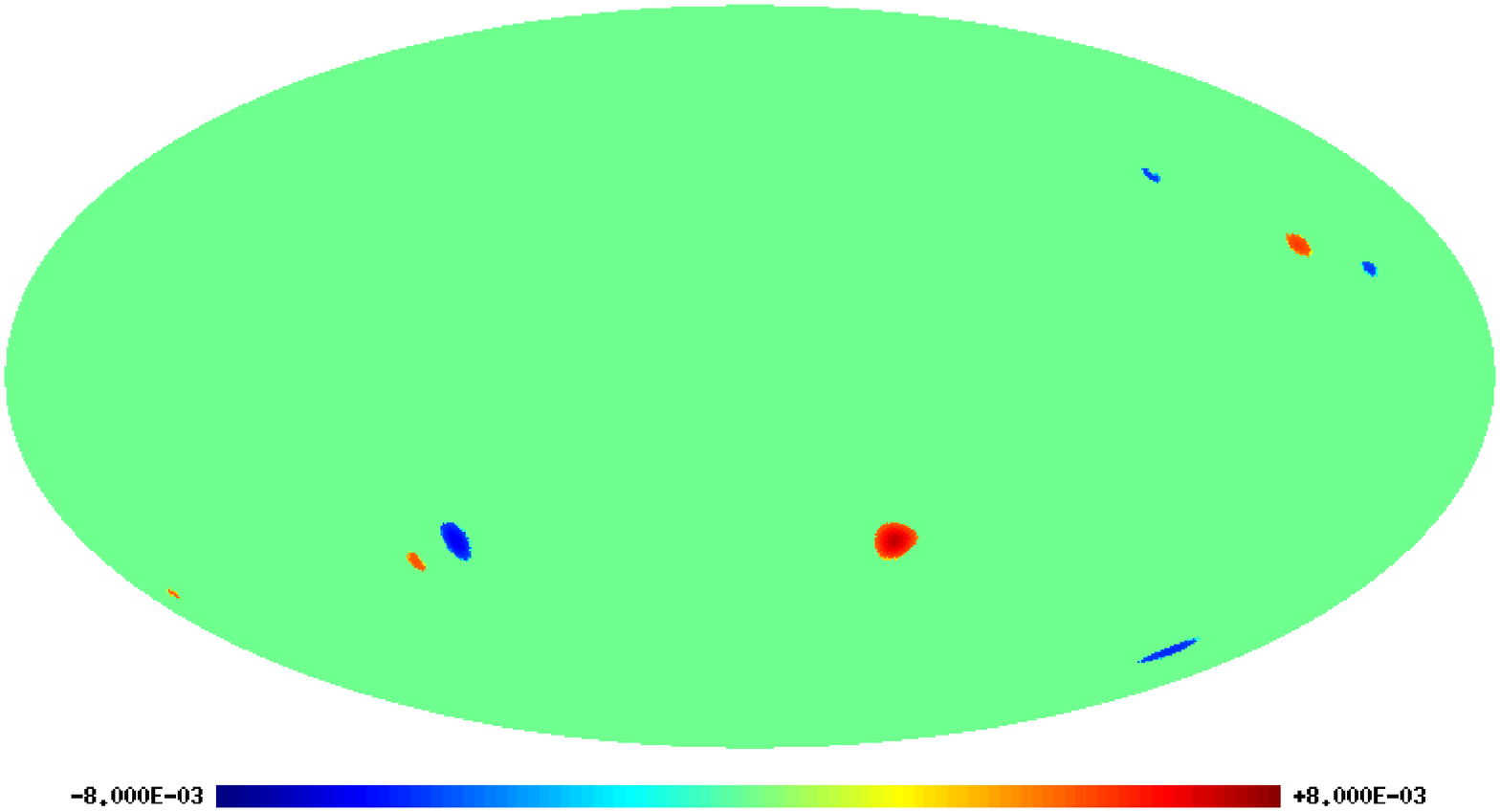}
      \end{minipage}\hspace{-3mm}
      \begin{minipage}[t]{25mm}
        \vspace{0pt}
        \frame{\includegraphics[bb= 570 50 700 120,width=20mm,clip]{figures/wmapcom_mbianchix1p55_mexhat000_thres_ia06_ig01}}
      \end{minipage}
    }
}
\mbox{
  \subfigure[Co-added \wmap\ map \mexhat\ wavelet coefficients ($\eccen=0.95$; $\scale_{10}=500\arcmin$; $\eulerc=108^\circ$)]
    { \hspace{5mm}
      \begin{minipage}[t]{55mm}
        \vspace{0pt}
        \includegraphics[width=\coeffplotwidth]{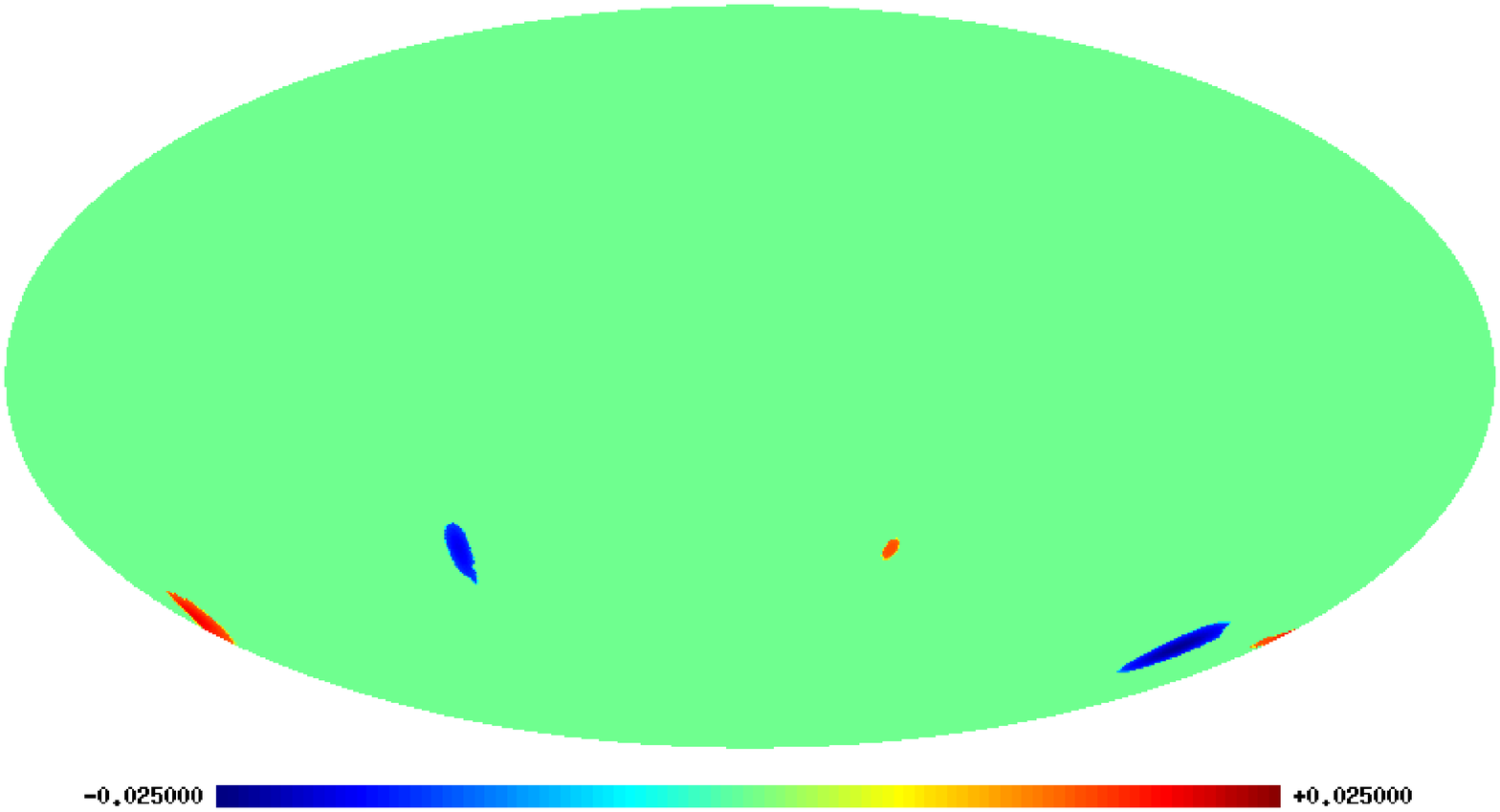}
      \end{minipage}\hspace{-3mm}
      \begin{minipage}[t]{25mm}
        \vspace{0pt}
        \frame{\includegraphics[bb= 570 50 700 120,width=20mm,clip]{figures/wmap_wcoeff_mexhat095_thres_ia10_ig05}}
      \end{minipage}
    }
  \hspace{5mm}
  \subfigure[Bianchi corrected co-added \wmap\ map \mexhat\ wavelet coefficients ($\eccen=0.95$; $\scale_{10}=500\arcmin$; $\eulerc=108^\circ$)]
    { \hspace{5mm}
      \begin{minipage}[t]{55mm}
        \vspace{0pt}
        \includegraphics[width=\coeffplotwidth]{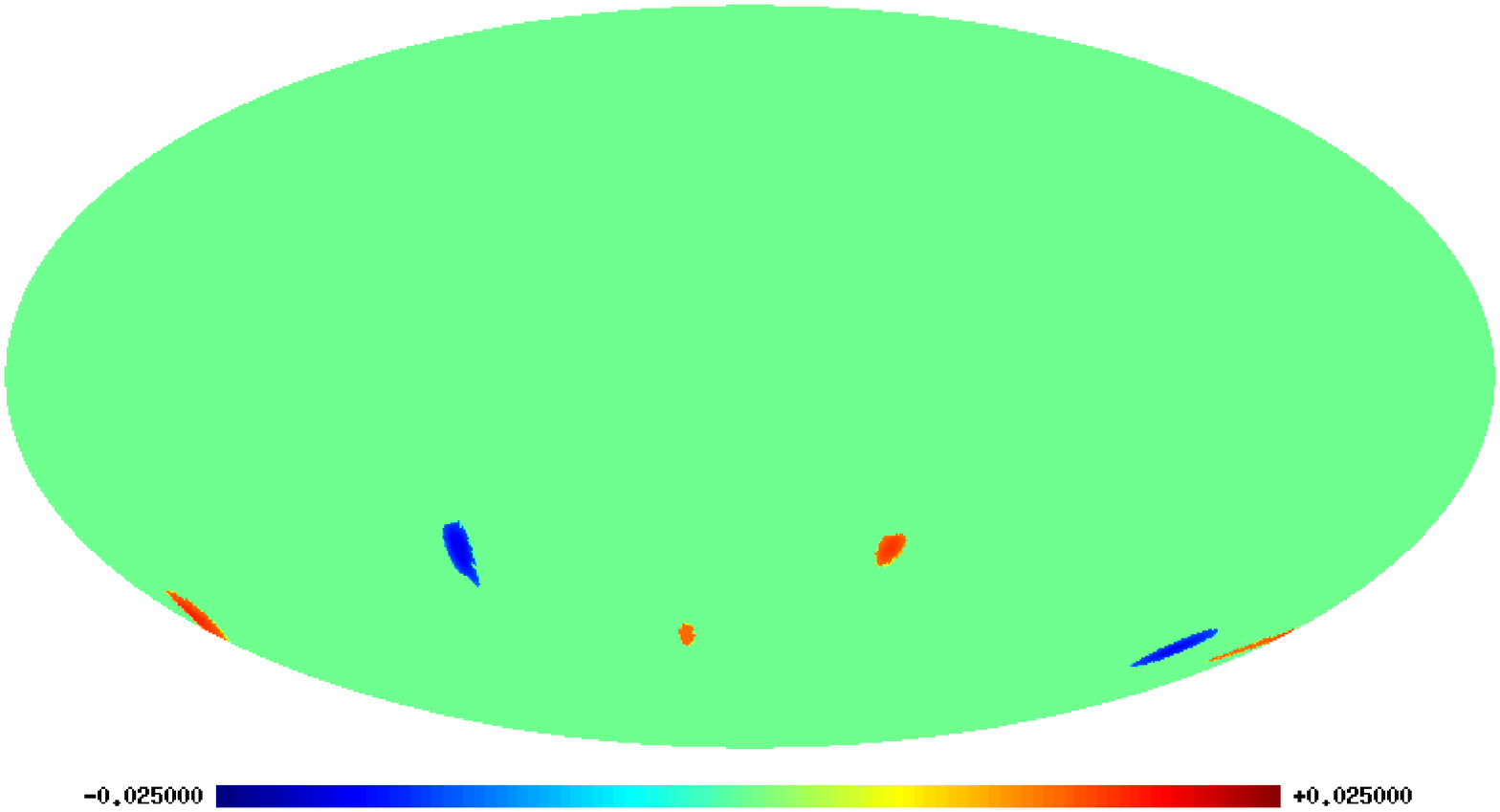}
      \end{minipage}\hspace{-3mm}
      \begin{minipage}[t]{25mm}
        \vspace{0pt}
        \frame{\includegraphics[bb= 570 50 700 120,width=20mm,clip]{figures/wmapcom_mbianchix1p55_mexhat095_thres_ia10_ig05}}
      \end{minipage}
    }
}
\caption{Thresholded spherical wavelet coefficients for the original and Bianchi corrected co-added \wmap\ map.
The inset figure in each panel shows a zoomed section (of equivalent size) around the cold spot at \spotloc.
The size and magnitude of this cold spot is reduced in the Bianchi corrected data.
Only those coefficient maps corresponding to the most significant kurtosis detections for the \mexhat\ wavelets are shown.  Other coefficient maps show no additional information to that presented in our previous work \citep{mcewen:2005a} (see text).
The corresponding wavelet coefficient maps for the \ilc\ map are not shown since they are almost identical to the coefficients of the co-added \wmap\ maps shown above.}
\label{fig:coeff}
\end{minipage}
\end{figure*}

\subsection{Gaussian plus Bianchi simulated \cmb\ map}

It addition to testing the \wmap\ data and Bianchi corrected versions of the data, we also consider a simulated map comprised of Gaussian \cmb\ fluctuations plus an embedded Bianchi component.  We use the same strategy to simulate the Gaussian component of the map that is used to create the Gaussian \cmb\ realisations used in the Monte Carlo analysis and add to it a scaled version of the Bianchi template that was fitted to the \wmap\ data by \jaffeshort.  The motivation is to see whether any localised regions in the map that contribute most strongly to non-Gaussianity coincide with any structure of the Bianchi template.

\begin{figure}
\centering
\subfigure[Gaussian plus Bianchi simulated map \mexhat\ wavelet coefficients ($\eccen=0.00$; $\scale_{6}=300\arcmin$)]{\includegraphics[width=\coeffplotwidth]{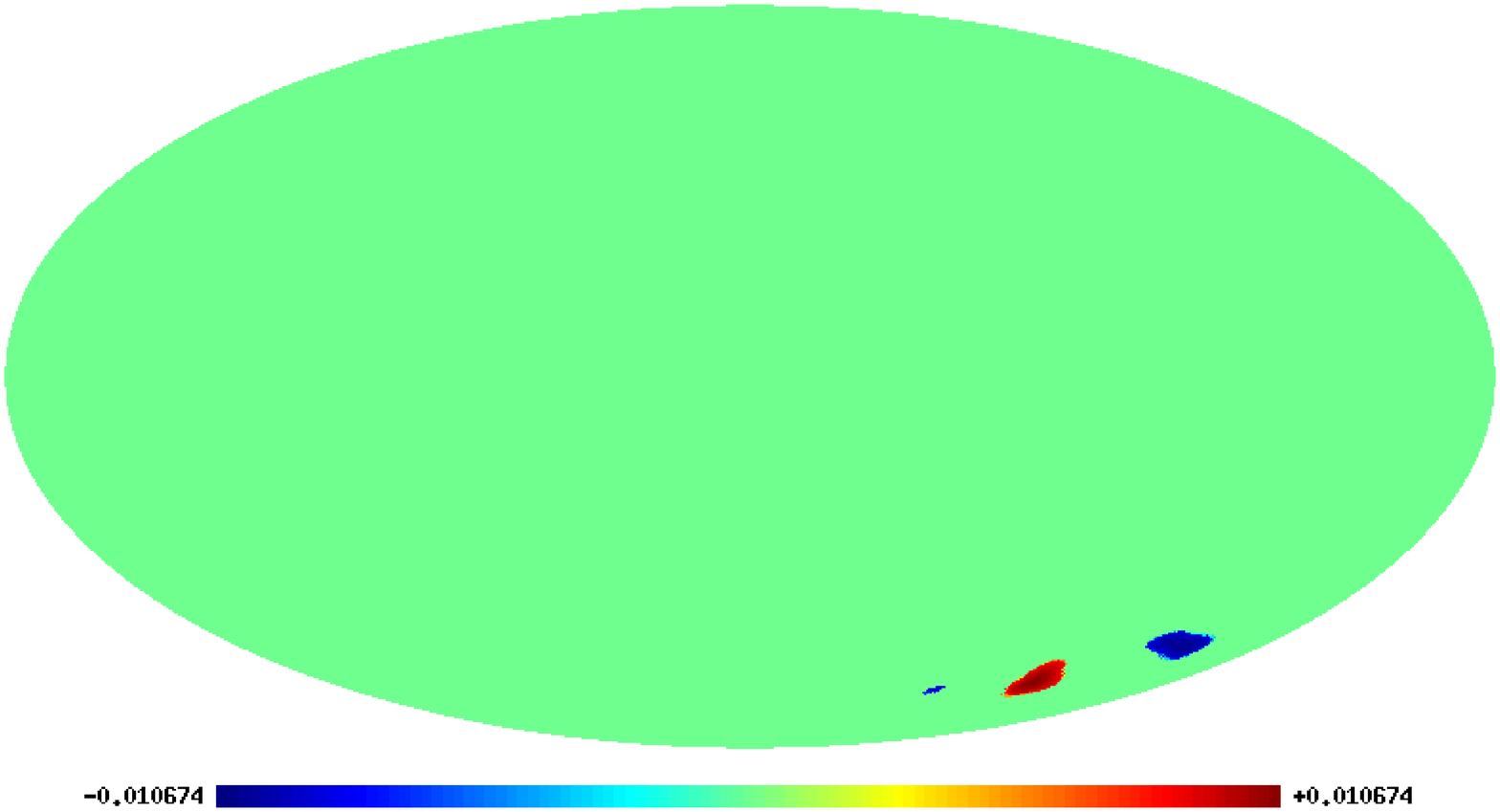}}
\subfigure[Gaussian plus Bianchi simulated map \mexhat\ wavelet coefficients ($\eccen=0.95$; $\scale_{10}=500\arcmin$; $\eulerc=108^\circ$)]{\includegraphics[width=\coeffplotwidth]{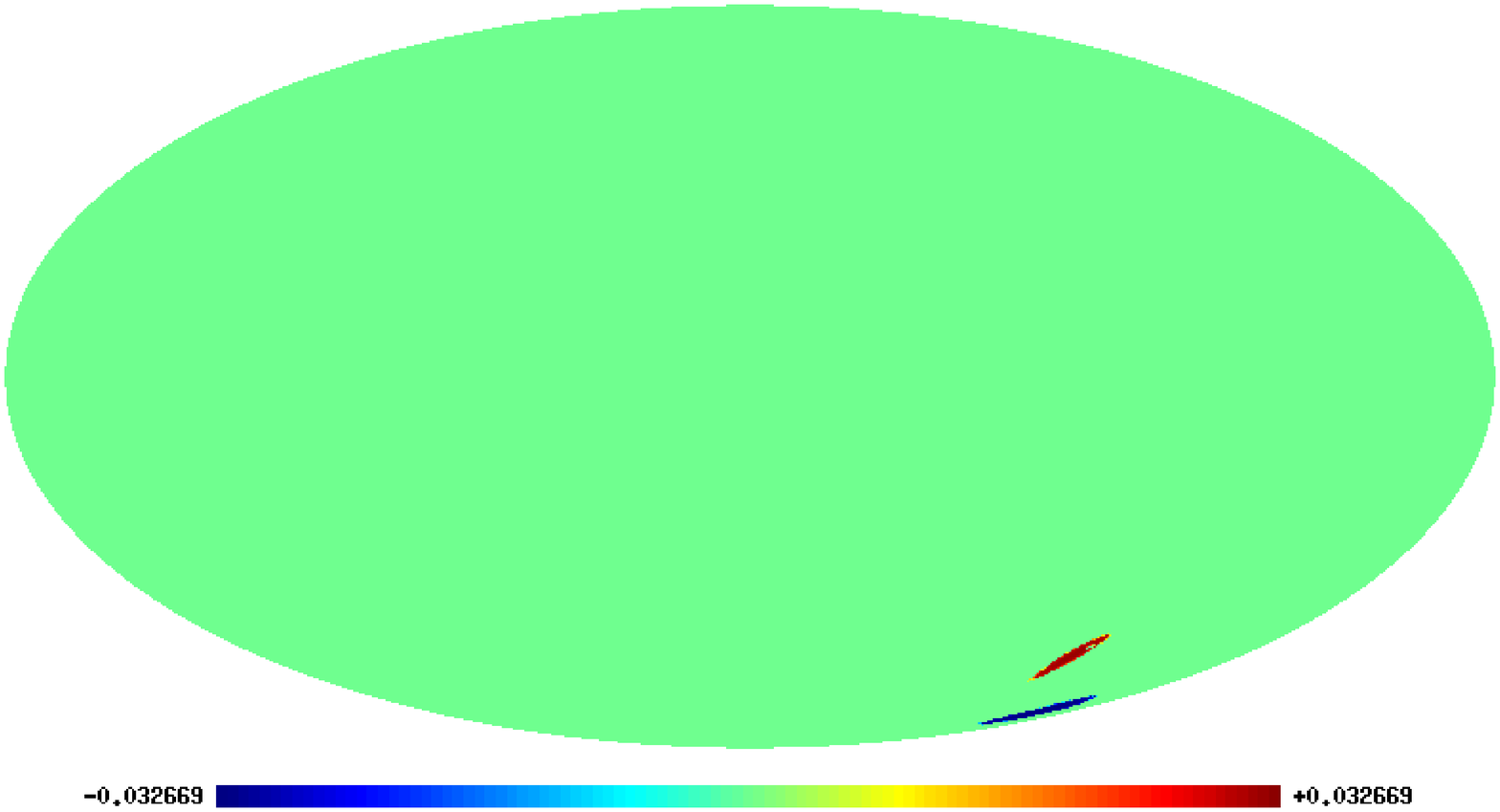}}
\caption{Thresholded \mexhat\ wavelet coefficients of the Guassian plus Bianchi simulated map.  The coefficient maps shown are flagged by a kurtosis detection of non-Gaussianity.  Notice how the \mexhat\ wavelets extract the intense regions near the centre of the Bianchi spiral, with the symmetric \mexhat\ wavelet ($\eccen=0.00$) extracting the symmetric structure and the elliptical \mexhat\ wavelet ($\eccen=0.95$) extracting the oriented structure.}
\label{fig:gsim_thres}
\end{figure}

Non-Gaussianity is detected at approximately the $3\sigma$ level in the kurtosis of the symmetric and elliptical \mexhat\ wavelet coefficients once the amplitude of the added Bianchi template is increased to approximately \mbox{$\bshear \sim 15 \times 10^{-10}$} (approximately four times the level of the Bianchi template fitted by \jaffeshort), corresponding to a vorticity of \mbox{$\bvort \sim 39 \times 10^{-10}$}.  No detections are made in any skewness statistics or with the \morlet\ wavelet.  The localised regions of the wavelet coefficient maps for which non-Gaussianity detections are made are shown in \fig{\ref{fig:gsim_thres}}.  The \mexhat\ wavelets extract the intense regions near the centre of the Bianchi spiral, with the symmetric \mexhat\ wavelet extracting the symmetric structure and the elliptical \mexhat\ wavelet extracting the oriented structure.  This experiment highlights the sensitivity to any Bianchi component of the \mexhat\ kurtosis statistics, and also the insensitivity of the \mexhat\ skewness statistics and \morlet\ wavelet statistics.  The high amplitude of the Bianchi component required to make a detection of non-Gaussianity suggests that some other source of non-Gaussianity may be present in the \wmap\ data, such as the cold spot at \spotloc, and that the Bianchi correction may act just to reduce this component.

%

\subsection{Foregrounds and systematics}

From the proceeding analysis it would appear that a Bianchi component is not responsible for the non-Gaussianity observed in the skewness of the spherical \morlet\ wavelet coefficients.  The question therefore remains: what is the source of this non-Gaussian signal?   We perform here a preliminary analysis to test whether unremoved foregrounds or \wmap\ systematics are responsible for the non-Gaussianity.

The coadded map analysed previously is constructed from a noise weighted sum of two Q-band maps observed at 40.7GHz, two V-band maps observed at 60.8GHz and four W-band maps observed at 93.5GHz.  To test for foregrounds or systematics we examine the skewness observed in the separate \wmap\ bands, and also in difference maps constructed from the individual bands.  In \fig{\ref{fig:stat_plot2}~(a)} the skewness of \morlet\ wavelet coefficients is shown for the individual band maps $\rm Q=Q1+Q1$, $\rm V=V1+V2$ and $\rm W=W1+W2+W3+W4$, and in \fig{\ref{fig:stat_plot2}~(b)} the skewness is shown for the difference maps $\rm V1-V2$, $\rm Q1-Q2$, $\rm W1-W4$ and $\rm W2-W3$ (the W-band difference maps have been chosen in this order to ensure that the beams of the maps compared are similar).  Note that the confidence regions shown in \fig{\ref{fig:stat_plot2}~(b)} correspond to the \wmap\ coadded map and not the difference maps.  It is computationally expensive to compute simulations and significance regions for the difference maps, thus one should only compare the skewness signal with that observed previously.
One would expect any detection of non-Gaussianity due to unremoved foregrounds to be frequency dependent.  The skewness signal we detect on scale $\scale_{11}$ is identical in all of the individual \wmap\ bands, hence it seems unlikely that foregrounds are responsible for the signal.
Moreover, since the skewness signal is present in all of the individual bands it would appear that the signal is not due to systematics present in a single \wmap\ channel.  The signal is also absent in the difference maps that are dominated by systematics and should be essentailly absent of \cmb\ and foregrounds.

From this preliminary analysis we may conclude that it is unlikely that foregound contamination or \wmap\ systematics are responsible for the highly significant non-Gaussianity detected with the spherical \morlet\ wavelet.  This analysis has also highlighted a possible systematic in the Q-band on scale $\scale_6$.  A more detailed analysis of this possible systematic and a deeper analysis of the cause of the non-Gaussianity detected with the \morlet\ wavelet is left for a separate piece of work.

\setlength{\statplotwidth}{65mm}

\begin{figure}
\centering
\subfigure[Individual \wmap\ band maps: WMAP {co\-added} (solid, blue, square); Q (solid, green, circle), V (dashed, blue, triangle); W (dashed, green diamond).]
  {\includegraphics[trim=0mm 0mm -3mm 0mm,clip,angle=-90,width=\statplotwidth]{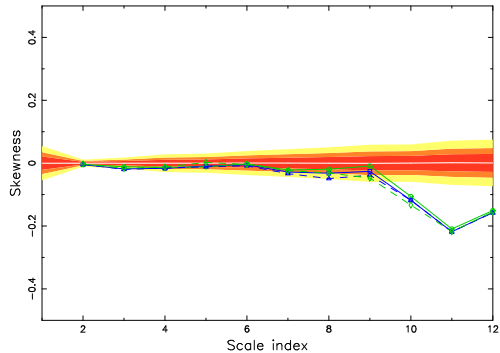}}
\subfigure[\wmap\ band difference maps: $\rm Q1-Q2$ (solid, blue, square); $\rm V1-V2$ (solid, green, circle); $\rm W1-W4$ (dashed, blue, triangle); $\rm W2-W3$ (dashed, green, diamond).]
  {\includegraphics[trim=0mm 0mm -3mm 0mm,clip,angle=-90,width=\statplotwidth]{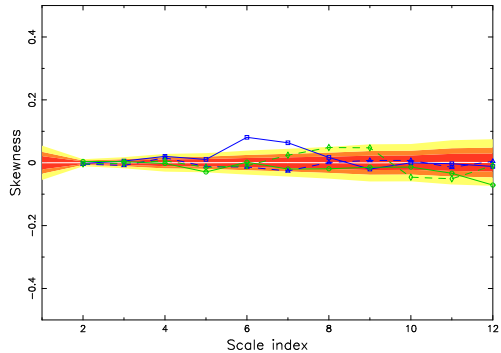}}
\caption{Skewness for individual and difference \wmap\ band maps -- \morlet\ \mbox{$\bmath{k}=\left( 10, 0 \right)^{T}$}.  Note that the strong non-Gaussianity detection made on scale $\scale_{11}$ is present in all of the individual band maps but is absent from all of the difference maps that should contain predominantly systematics.  The confidence regions shown in these plots are for the \wmap\ coadded map (see comment text).}  
\label{fig:stat_plot2}
\end{figure}


\section{Conclusions}
\label{sec:conclusions}

We have investigated the effect of correcting the \wmap\ data for a Bianchi type VII$_{\rm h}$ template on our previous detections of non-Gaussianity made with directional spherical wavelets \citep{mcewen:2005a}.
The best-fit Bianchi template was simulated with the parameters determined by \jaffeshort\ using the latest shear and vorticity estimates (\jaffeshort; private communication).
We subsequently used this best-fit Bianchi template to `correct' the \wmap\ data, and then repeated our wavelet analysis to probe for deviations from Gaussianity in the corrected data.

The deviations from Gaussianity observed in the kurtosis of spherical wavelet coefficients disappears after correcting for the Bianchi component, whereas the deviations from Gaussianity observed in skewness statistics are not affected.
The highly significant detection of non-Gaussianity previously made in the skewness of \morlet\ wavelet coefficients remains unchanged at 98\% using the extremely conservative method to compute the significance outlined in \sectn{\ref{sec:stat_sig}}.
The $\chi^2$ tests also performed indicate that the Bianchi corrected data still deviates from Gaussianity when all test statistics are considered in aggregate.
Since only the skewness-flagged detections of non-Gaussianity made with the \mexhat\ wavelet remain, but the kurtosis ones are removed, the overall significance of \mexhat\ wavelet $\chi^2$ tests are reduced.
There was no original detection of kurtosis in the \morlet\ wavelet coefficients, thus the significance of the $\chi^2$ remains unchanged for this wavelet.
Finally, note that one would expect the skewness statistics to remain unaffected by a Bianchi component (or equivalently the removal of such a component) since the distribution of the pixel values of a Bianchi component is itself not skewed, whereas a similar statement cannot be made for the kurtosis.

Regions that contribute most strongly to the non-Gaussianity detections have been localised.  The skewness-flagged regions of the Bianchi corrected data do not differ significantly from those regions previously found in \citet{mcewen:2005a}.  One would expect this result: if these regions are indeed the source of non-Gaussianity, and the non-Gaussianity is not removed, then when most likely contributions to non-Gaussianity are again localised the regions should remain.  The kurtosis-flagged regions localised with the \mexhat\ wavelets are not markedly altered by correcting for the Bianchi template, however the size and magnitude of the cold spot at Galactic coordinates \spotloc\ is significantly reduced.
\citet{cruz:2005} claim that it is solely this cold spot that is responsible for the kurtosis detections of non-Gaussianity made with \mexhat\ wavelets, thus the reduction of this cold spot when correcting for the Bianchi template may explain the elimination of kurtosis in the Bianchi corrected maps.

After correcting the \wmap\ data for the best-fit \bianchiviih\ template, the data still exhibits significant deviations from Gaussianity, as highlighted by the skewness of spherical wavelet coefficients.  A preliminary analysis of foreground contamination and \wmap\ systematics indicates that these factors are also not responsible for the non-Gaussianity.  A deeper investigation into the source of the non-Gaussianity detected is required to ascertain whether the signal is of cosmological origin, in which case it would provide evidence for non-standard cosmological models.
Bianchi models that exhibit a small universal shear and rotation are an important, alternative cosmology that warrant investigation
and, as we have seen, can account for some detections of non-Gaussian signals.
However, the current analysis is only phenomenological since revisions are required to update the Bianchi-induced temperature fluctuations calculated by \citet{barrow:1985} for a more modern setting.
Nevertheless, such an analysis constitutes the necessary first steps towards examining and raising the awareness of anisotropic cosmological models.


\section*{Acknowledgements}

We thank Tess Jaffe and Anthony Banday for useful discussions on their simulation of \cmb\ temperature fluctuations induced in \bianchiviih\ models.
JDM thanks the Association of Commonwealth
Universities and the Cambridge Commonwealth Trust for the 
support of a Commonwealth (Cambridge) Scholarship.
DJM is supported by PPARC.
Some of the results in this paper have been derived using the
\healpix\ package \citep{gorski:2005}.
We acknowledge the use of the Legacy Archive for Microwave Background
Data Analysis (\lambdaarch).  Support for \lambdaarch\ is provided by
the NASA Office of Space Science.



\appendix
\section{Bianchi type VII$_{\rm \lowercase{h}}$ representations}
\label{sec:appn_bianchi}

We do not attempt to describe or solve the geodesic equations resulting from the \bianchiviih\ model, but rather refer to reader to \citet{barrow:1985} for details.  Enough detail on the induced temperature fluctuations is presented here so that the interested reader may re-produce our simulated maps.

Analytic forms are given to compute the Bianchi-induced temperature fluctuations directly in either real or harmonic space.
We correct some errors (most likely typographical) in \citet{barrow:1985} in the analytic form of the harmonic space representation.
Different representations are preferred for different applications, hence the ability to compute directly in the space required is obviously beneficial.
The efficiency of computing in either space is not markedly different, however computing the harmonic coefficients directly does have some advantages.
For instance, the $\thetaalm$ resolution required to compute numerically some integrals
(specifically, \eqn{\ref{eqn:ila}} and \eqn{\ref{eqn:ilb}})
is independent of the resolution of the real space map, and typically the integrals may be computed accurately for a relatively low \thetaalm\ resolution.
Moreover, the rotation may be performed considerably more efficiently and accurately%
\footnote{A finite point set on the sphere that is invariant under rotations does not
exist, thus arbitrary rotations cannot be performed exactly in real space on a
pixelised sphere.} %
in harmonic space.
Finally, the Bianchi-induced fluctuations have a low band-limit, thus harmonic coefficients need only be computed for a relatively low \lmax.

\subsection{Real space representation}

The \bianchiviih -induced temperature fluctuations are parameterised by
the total energy density \omegat, the redshift of recombination \ze\ and the \bianchi\ shear \bshear, handedness \bhand\ and \bx\ parameters, where \hub\ is the Hubble parameter.  The handedness parameter takes the values $\bhand=+1$ and
$\bhand=-1$ for maps with right and left handed spirals respectively.
Physically, \bx\ is related to the characteristic wavelength over which the principle axes of shear and rotation change orientation; hence \bx\ defines the `spiralness' of the resultant temperature fluctuations.  The \bh\ parameter of the \bianchiviih\ model is related to \bx\ and \omegat\ by
\begin{equation}
\bx=\sqrt{\frac{\bh}{1-\omegat}}
\spcend .
\end{equation}
The vorticity of a \bianchiviih\ model is related to the other parameters by
\begin{equation}
\label{eqn:vort}
\bvort= \frac{\sqrt{2} (1+\bh)^{1/2} (1+9\bh)^{1/2}}{6\bx^2\omegat} \bshear
\spcend .
\end{equation}

The temperature fluctuations induced in the Bianchi type VII$_{\rm h}$ model are given by
\begin{eqnarray}
\frac{\dtemp}{\temp} (\thetaph, \phiph) &=&
\bshear \{ [\ba(\thetaph) + \bb(\thetaph) ] \sin(\phiph) \nonumber \\
&& \mbox{} + 
\bhand \,
[\bb(\thetaph) - \ba(\thetaph) ] \cos(\phiph) \}
\spcend ,
\label{eqn:temp_real}
\end{eqnarray}
where \thetaph\ and \phiph\ specify the direction of the incoming photon, which is related to the observing angles by $\thetaob=\pi-\thetaph$ and 
$\phiob=\pi+\phiph$.
$\ba(\thetaph)$ and $\bb(\thetaph)$ are defined by
\begin{eqnarray}
\ba(\thetaph) &=& \bcone  \sin(\thetaph) \nonumber \\
&& \mbox{} -
\bcone \bctwo(\thetaph)\left\{ \cos\left[\bps(\bt_E,\thetaph)\right]
  - 3h^{1/2} \sin\left[\bps(\bt_E, \thetaph)\right] \right\} 
\nonumber \\
&& \mbox{}
+ \bcthree \int_{\bt_E}^{\bt_0}
\frac{\bs(\bt, \thetaph)\left[1-\bs^2(\bt, \thetaph)\right] \sin\left[\bps(\bt, \thetaph)\right] {\rm d}\bt }
{\left[1+\bs^2(\bt, \thetaph)\right]^2 \sinh^4(\bh^{1/2}\bt/2) }
\label{eqn:atheta}
\end{eqnarray}
and
\begin{eqnarray}
\bb(\thetaph) &=& 3\bh^{1/2} \bcone  \sin(\thetaph) \nonumber \\
&& \mbox{} -
\bcone \bctwo(\thetaph)\left\{ \sin\left[\bps(\bt_E,\thetaph)\right]
  + 3h^{1/2} \cos\left[\bps(\bt_E, \thetaph)\right] \right\} 
\nonumber \\
&& \mbox{}
- \bcthree \int_{\bt_E}^{\bt_0}
\frac{\bs(\bt, \thetaph)\left[1-\bs^2(\bt, \thetaph)\right] \cos\left[\bps(\bt, \thetaph)\right] {\rm d}\bt }
{\left[1+\bs^2(\bt, \thetaph)\right]^2 \sinh^4(\bh^{1/2}\bt/2) }
\spcend ,
\label{eqn:btheta}
\end{eqnarray}
where we integrate over the photon path in conformal time \bt\ from
photon emission (last scattering surface)
\begin{equation}
\bt_E = 2 \bh^{-1/2} \sinh^{-1}\left[ \left(\frac{\omegat^{-1}-1}{1+\ze}\right)^{1/2} \right]
\spcend ,
\end{equation}
to observation
\begin{equation}
\bt_0 = 2 \bh^{-1/2} \sinh^{-1}\left[ (\omegat^{-1}-1)^{1/2} \right]
\spcend .
\end{equation}
The terms \bcone, $\bctwo(\thetaph)$, \bcthree\ are defined by
\begin{equation}
\bcone = (3 \omegat \bx)^{-1}  
\spcend ,
\end{equation}
\begin{equation}
\bctwo(\thetaph) = \frac{2 \, \bs(\bt_E,\thetaph) \, (1+\ze)}{1+\bs^2(\bt_E,\thetaph)}
\spcend ,
\end{equation}
and
\begin{equation}
\bcthree = 4 \bh^{1/2} (1-\omegat)^{3/2} \omegat^{-2}
\spcend .
\end{equation}
The functions $\bs(\bt, \thetaph)$ and $\bps(\bt, \thetaph)$ are defined by
\begin{equation}
\bs(\bt, \thetaph) =  \exp{ - h^{1/2} (\bt - \bt_0) } \tan(\thetaph/2)
\end{equation}
and
\begin{eqnarray}
\bpsi(\bt, \thetaph) &=& (\bt-\bt_0) - \bh^{-1/2}
\ln \left[ \sin^2(\thetaph/2) \right. \nonumber \\
&&\left. \mbox{}+ \exp{ 2 \bh^{1/2} (\bt-\bt_0)} \cos^2(\thetaph/2) \right]
\spcend .
\end{eqnarray}
%
%
At a sacrifice to notational simplicity, but, we hope, to add clarity, we have made the dependence on all variables that are not intrinsic Bianchi parameters explicit in all functions.

The induced swirl pattern described by these equations is centred on the south pole.  Obviously, the centre of the swirl may by located at any position in our coordinate system, thus a final rotation is required to fully describe any possible realisation of the Bianchi-induced temperature fluctuations.  This introduces three additional parameters, the three Euler angles that describe the rotation $(\eulera,\eulerb,\eulerc)$ (for which we adopt the active $zyz$-convention, where functions rather than axes are rotated).


\subsection{Harmonic space representation}

We derive the analytic form of the spherical harmonic coefficients of the \bianchi -induced temperature fluctuations in this section.
%
Due to the low frequency structure of the Bianchi temperature fluctuations, both globally and azimuthally, the band-limit of harmonic coefficients is low in both \el\ and $m$.  Thus, computing simulated temperature fluctuations may be performed efficiently and, as we explain, more accurately in harmonic space.

We decompose the Bianchi-induced temperature fluctuations into a sum of spherical harmonics
\begin{equation}
\label{eqn:temp_harmexp}
\frac{\dtemp}{\temp} (\thetaalm, \phialm) =
\sum_{\el=1}^{\infty}
\sum_{\m=-\el}^{\el}
\almi
\sh{\el}{\m}{\thetaalm,\phialm}
\spcend ,
\end{equation}
where the harmonic coefficients are given by the usual projection on to each spherical harmonic basis function
\begin{equation}
\label{eqn:almdef}
\almi = \int_\sphere
\frac{\dtemp}{\temp} (\thetaalm, \phialm) \, \shc{\el}{\m}{\thetaalm,\phialm}
\,
\domega
\spcend ,
\end{equation}
and $\domega = \sin\thetaalm \dx \thetaalm \dx \phialm$ is the usual rotation invariant measure on the two-sphere \sphere.  In practice the outer summation of \eqn{\ref{eqn:temp_harmexp}} is truncated to \lmax\ terms.
We adopt the Condon-Shortley phase convention, where the normalised spherical harmonic are defined by
\begin{equation}
\sh{\el}{\m}{\thetaalm, \phialm} = (-1)^\m \sqrt{\frac{2\el+1}{4\pi}
\frac{(\el-\m)!}{(\el+\m)!}} \,
\aleg{\el}{\m}{\cos\thetaalm} \,
\exp{\img \m \phialm}
\spcend ,
\end{equation}
where $\aleg{\el}{\m}{x}$ are the associated Legendre functions.
If we adopt the photon, rather than observing angles for the temperature fluctuations we simply introduce a factor of $(-1)^{\el+\m}$:
\begin{equation}
\almitilde = \int_\sphere
\frac{\dtemp}{\temp} (\thetaph, \phiph) \, \shc{\el}{\m}{\thetaob,\phiob}
\, \domega_{\rm ob}
= (-1)^{\el+\m} \almi
\spcend .
\end{equation}
Substituting the expression for the temperature fluctuations given by \eqn{\ref{eqn:temp_real}} into \eqn{\ref{eqn:almdef}} yields
\begin{equation}
\label{eqn:temp_alm}
\almi=
\bshear \pi [\m \bhand (\bi{\ba}{\el} - \bi{\bb}{\el})
+ \img (\bi{\ba}{\el} + \bi{\bb}{\el})]
\kron_{\m,\pm 1}
\spcend ,
\end{equation}
where $\kron_{i,j}$ is the Kronecker delta,
\begin{equation}
\label{eqn:ila}
\bi{\ba}{\el} = \sqrt{\frac{2\el+1}{4\pi\el(\el+1)}} \int_0^\pi
\ba(\thetaalm) \aleg{\el}{1}{\cos\thetaalm} \sin\thetaalm \dx\thetaalm
\end{equation}
and
\begin{equation}
\label{eqn:ilb}
\bi{\bb}{\el} = \sqrt{\frac{2\el+1}{4\pi\el(\el+1)}} \int_0^\pi
\bb(\thetaalm) \aleg{\el}{1}{\cos\thetaalm} \sin\thetaalm \dx\thetaalm
\spcend .
\end{equation}
In the derivation of \eqn{\ref{eqn:temp_alm}} we have made use of 
the following relations:
\begin{equation}
\int_0^{2\pi} \exp{-\img \m \phialm} \sin\phialm \dx \phialm =
-\img\m\pi \,
\kron_{\m,\pm 1}
\spcend ,
\end{equation}
\begin{equation}
\int_0^{2\pi} \exp{-\img \m \phialm} \cos\phialm \dx \phialm =
\pi \,
\kron_{\m,\pm 1}
\end{equation}
and
\begin{equation}
\aleg{\el}{-\m}{x} = (-1)^\m \frac{(\el-\m)!}{(\el+\m)!} \aleg{\el}{\m}{x}
\spcend .
\end{equation}
Notice that the minimal azimuthal structure of the induced temperature fluctuations ensures that the harmonic coefficients are non-zero only for $\m=\pm1$.
For completeness, we also state the analytic form for the power spectrum of the Bianchi-induced fluctuations:
\begin{equation}
C_\el = \frac{4\pi^2}{2\el+1}
\bshear^2
\left[ (\bi{\ba}{\el})^2 + (\bi{\bb}{\el})^2 \right]
\spcend .
\end{equation}

The swirl pattern of the induced temperature fluctuations is again centred on the south pole in the expression given by \eqn{\ref{eqn:temp_alm}}.  We may perform the required rotation directly in harmonic space, noting that the spherical harmonics of a rotated function are related to the harmonics of the original function by \citep{risbo:1996}
\begin{equation}
\almi^{\rm rot} = \sum_{\m\p=-\el}^{\el} D^\el_{\m\m\p}(\eulers) \, \almpi
\spcend ,
\end{equation}
where $D^\el_{\m,\m\p}$  are the Wigner D-matrices.
The harmonic coefficients of the rotated Bianchi-induced temperature fluctuation map simply reduce to
\begin{equation}
\almi^{\rm rot} = D^\el_{\m,-1}(\eulers) \, \alm_{\el,-1}
+ D^\el_{\m,+1}(\eulers) \, \alm_{\el,+1}
\spcend ,
\end{equation}
since the harmonic coefficients of the original map are non-zero only for $\m=\pm1$.


\label{lastpage}

\end{document}